\documentclass[draftclsnofoot,onecolumn,12pt]{IEEEtran}

\usepackage{cite,graphicx,amsmath,amssymb,dsfont}
\usepackage{color,psfrag}
\usepackage{booktabs}
\usepackage{citesort}
\usepackage{amsthm}
\usepackage{stfloats}
\usepackage{times}
\usepackage{relsize}
\usepackage{tabularx}
\usepackage{epstopdf}
\usepackage{multirow}
\usepackage{subcaption}
\usepackage[font=scriptsize]{caption}
\makeatletter
\renewcommand{\maketag@@@}[1]{\hbox{\m@th\normalsize\normalfont#1}}%
\makeatother

\newtheorem{theorem}{Theorem}

\newtheorem{corollary}{\bf Corollary}
\newtheorem{remark}{\bf Remark}

\newtheorem{assumption}{\bf Assumption}

\makeatletter
\newcommand{\figcaption}{\def\@captype{figure}\caption}
\newcommand{\tabcaption}{\def\@captype{table}\caption}
\makeatother

\newcommand{\bs}{\boldsymbol}

\newcommand{\bbE}{\mathbb{E}}

\newcommand{\bbP}{\mathbb{P}}
\newcommand{\bbR}{\mathbb{R}}

\newcommand{\bg}{\mathbf{g}}

\newcommand{\mbf}{\mathbf{f}}

\newcommand{\bh}{\mathbf{h}}

\newcommand{\bhg}{\mathbf{\hat g}}

\newcommand{\btg}{\mathbf{\tilde {g}}}

\newcommand{\nsp}{\negthickspace}
\newcommand{\nmsp}{\negmedspace}

\newcommand{\papertitle}{Heterogeneous Cellular Networks with LoS and NLoS Transmissions---The Role of Massive MIMO and Small Cells}

\begin{document}


\title{\papertitle}

\author{Qi Zhang, Howard H. Yang, Tony Q. S. Quek, and Jemin Lee
\thanks{Qi Zhang, Howard H. Yang and Tony Q. S. Quek are with Singapore University of Technology and Design, Singapore 487372, Singapore (email: qi$\_$zhang@sutd.edu.sg; eehowardh@gmail.com; tonyquek@sutd.edu.sg).}
\thanks{Jemin Lee is with Daegu Gyeongbuk Institute of Science and Technology, Daegu 42988, Korea (email: jmnlee@dgist.ac.kr).}}

\maketitle

\begin{abstract}
We develop a framework for downlink heterogeneous cellular networks with line-of-sight (LoS) and non-line-of-sight (NLoS) transmissions. Using stochastic geometry, we derive tight approximation of achievable downlink rate that enables us to compare the performance between densifying small cells and expanding BS antenna arrays. Interestingly, we find that adding small cells into the network improves the achievable rate much faster than expanding antenna arrays at the macro BS. However, when the small cell density exceeds a critical threshold, the spacial densification will lose its benefits and further impair the network capacity. To this end, we present the optimal small cell density that maximizes the rate as practical deployment guidance. In contrast, expanding macro BS antenna array can always benefit the capacity until an upper bound caused by pilot contamination, and this bound also surpasses the peak rate obtained from deployment of small cells. Furthermore, we find that allocating part of antennas to distributed small cell BSs works better than centralizing all antennas at the macro BS, and the optimal allocation proportion is also given for practical configuration reference. In summary, this work provides a further understanding on how to leverage small cells and massive MIMO in future heterogeneous cellular networks deployment.
\end{abstract}
\begin{IEEEkeywords}
Heterogeneous cellular networks, LoS/NLoS, massive MIMO, small cells, stochastic geometry
\end{IEEEkeywords}
\newpage



\section{Introduction}\label{sec: introduction}
The mobile data traffic has been doubling each year during the last few years and the wireless industry is preparing a $1000$-fold increase in data demands expected in this decade. To deal with this challenge, the fifth generation (5G) communications system has come at the forefront of wireless communications theoretical research
\cite{jungnickel14}. Two main approaches in 5G are massive antennas and dense deployments of access points, which lead to the massive multiple-input multiple-output (MIMO) and small cell techniques \cite{hwang13}.

Massive MIMO employs hundreds of antenna elements at the base station (BS) to serve tens of users simultaneously at the same time-frequency resource \cite{rusek13,larsson14,pitarokoilis12,wagner12}.
The large size of transmit antenna array not only significantly increases the capacity through excessive spatial dimensions \cite{marzetta10,jose11,yin13,Hoydis13,fernandes13,qi14,qi15,Ngo11},
but also averages out the effect of fast channel fading and
provides extremely sharp beamforming concentrated into small areas \cite{rusek13,larsson14}. Aside from these, the huge degrees-of-freedom offered by massive MIMO also reduce the transmit power \cite{Ngo11}. Nevertheless, due to the finite channel coherence time, performance of Massive MIMO is mainly limited by pilot contamination which arises from pilot reuse among adjacent cells \cite{marzetta10,jose11,yin13}.

On the other track, small cell improves the system capacity by densely deploying low-power access points into the traditional high-power macro cells \cite{quek13,hwang13}. In this fashion, distance between transmitter and receiver can be significantly reduced which results in remarkably enhanced rate gains. As small cells do not always have direct links to the macro BS, they can be intelligently deployed in accordance to the traffic demand without much cost on the fiber usage and real estate. However, the performance of small cell is mainly affected by the additional inter-cell interference induced from massive transmitting nodes
\cite{jungnickel14,wildemeersch13,lopez11}.

As both massive MIMO and small cells have attractive attribute in capacity enhancement, it's natural to wonder which one performs better under which scenarios. Comparison of massive MIMO and small cell from special and energy efficiency is addressed in \cite{nguyen16,liu14,bjornson13}. However, \cite{nguyen16,liu14} model the small cell system as a one-tier network with dense BSs, which is unreasonable since small cell is designed to offload heavy traffic from macro BS. In \cite{bjornson13}, a two-tier architecture including massive-antenna macro BS and small-cell access points is explored, but it only considers the single-cell scenario and ignores the randomness of BSs' locations. The flexible add-on small cells and further influence on user associations make it necessary to capture BSs' locations into analysis. Therefore, a reasonable framework to compare massive MIMO and small cell should be a heterogeneous cellular network (HCN) containing different types of multi-antenna and randomly located BSs. Stochastic geometry provides a useful tool to describe spatial distribution of BS sites and traffic flow \cite{andrews11}. Based on this, lots of work have focused on the HCN with Poisson point process (PPP) distributed BSs \cite{dhillon12,jo12,dhillon13,gupta14,yang15}. A general multi-tier framework is proposed in \cite{dhillon12}, and the coverage probability with flexible biased cell association is analyzed in \cite{jo12}. On top of these, further extensions to multi-antenna transmission are presented in \cite{dhillon13,gupta14,yang15}, which is much more challenging than the single-antenna scenario. However, none of these works can be applied directly on massive MIMO systems, since they do not take into account channel estimation and the further effect of pilot contamination, which is the main limiting factor of massive MIMO. Moreover, all these works adopt the single-slope path loss model with only non-line-of-sight (NLoS) transmissions, which is not fit for dense small cell networks since the short propagation distance results in more line-of-sight (LoS) transmissions \cite{zhang15} and multi-slope path loss affects the benefit of densification remarkably \cite{ding16}.

In this paper, we propose a framework for downlink HCN where the user location as well as the deployment of multi-antenna macro and small BSs are modeled as independent PPPs.
In this network, users are flexibly associated with the strongest BS, and each BS simultaneously serves multiple users associated with it on the same time-frequency resource block. Further, the signal propagation experiences a path loss model differentiating LoS and NLoS transmissions, and the channel state information at BSs is acquired through uplink training. Using stochastic geometry tools, we quantify the rate performance in a general setting that accounts for the interference affecting both the channel estimation and data transmission phases.
Our main contributions are summarized as follows:
\begin{itemize}
  \item We propose a general framework for the analysis of downlink HCN which consists of PPP distributed macro and small cell BSs with multiple antennas. Our analysis captures the essential keypoints of both massive MIMO and small cells, including LoS/NLoS transmission, BS deployment density, imperfect channel estimation, and random network topology.
  \item Based on our analytical results, we compare the system performance between densifying small cells and expanding macro BS antennas. It is found that adding small cells into the network can improve the achievable rate much faster than expanding antenna arrays at the macro BS. However, when the small cell density exceeds a critical threshold, the spacial densification will stop benefiting and further impair the network capacity. In contrast, the achievable rate always increases with growing antenna size and saturates to an upper bound caused by pilot contamination. This upper bound is also larger than the peak rate obtained from deployment of small cells.
  \item We provide the optimal small cell density that maximizes the achievable downlink rate, which can be used as a rule-of-thumb for practical small cell deployment. To fully exploit the degrees-of-freedom offered by the large antenna array, the optimal bias with different antenna numbers is provided as guidance for practical massive MIMO configuration.
  \item We also investigate the effect of distributed and centralized antennas. It is found that to attain higher data rate with fixed budget of antenna number, taking certain amount of antennas into distributed small cell BSs is more beneficial than centralizing all antennas at the macro BS. The optimal antenna allocation proportion is also presented as a reference for practical antenna configuration while combining with the hardware constraint.
\end{itemize}

The remainder of this paper is organized as follows. Section \ref{sec: system model} introduces the HCN framework, LoS/NLoS transmission model, and the user association policy. In Section \ref{sec: achievable downlink rate}, we derive a tight approximation of the achievable downlink rate accounting for channel estimation with uplink training. In Section \ref{sec: numerical results}, we provide numerical results to validate the analytical results and further study the performance of the HCN. Finally, Section \ref{sec: conclusion} summarizes the main results of this paper.

{\em Notation}---Throughout the paper, vectors are expressed in lowercase boldface letters while matrices are denoted by uppercase boldface letters. We use ${{\bf{X}}^H}$ to denote the conjugate-transpose of $\bf{X}$, and use $[{\bf{X}}]_{ij}$ to denote the ($i,j$)-th entry of $\bf{X}$.
Finally, $\mathds{1}(e)$ is the indicator function for logic $e$, $\mathbb{E}\left\{  \cdot  \right\}$ is the expectation operator and $\left\| {\, \cdot \,} \right\|$ is the Euclidean norm.

\section{System model}\label{sec: system model}
In this section, we introduce the network topology, the propagation model, and the user association policy. The main notations used throughout the paper are summarized in Table \ref{table: notation}.

\begin{table}
\caption{Notation Summary} \label{table: notation}
\begin{center}
\begin{tabular}{c p{13cm} }
\hline
 {\bf Notation} & {\hspace{2.5cm}}{\bf Definition}
\\
\midrule
\hline
$M_{\tt m}; M_{\tt s}$ & Number of antennas at each MBS and SCB \\
$\Phi_{\tt m}; \Phi_{\tt s}; \Phi_{\tt u}$ & PPPs modeling locations of MBSs, SCBs and users   \\
$\lambda_{\tt m}; \lambda_{\tt s}; \lambda_{\tt u}$ & Spatial densities of MBSs, SCBs and users \\
$P_{\tt m}; P_{\tt s}$ & Transmit power of each MBS and SCB \\
$B; \bbR$ & Bias factor; achievable downlink rate of the typical user \\
$N; U_{nl}$ & Number of users scheduled by each MBS; user $n$ scheduled by MBS $l$\\
$\bbP_{\tt m}^{\tt L}(\cdot); \bbP_{\tt s}^{\tt L}(\cdot)$ & LoS probability function of MBS and SCB \\
$L^{\tt L}; L^{\tt NL}$ & Path loss at a reference distance $1$ for LoS and NLoS \\
$\alpha^{\tt L}; \alpha^{\tt NL}$ &Path loss exponent for LoS and NLoS \\
$h_{minl}^{(\tt m)}; h_{mjnl}^{(\tt s)}$ & Small-scale fading from $U_{nl}$ to the $m$-th antenna of MBS $i$ and SCB $j$ \\
$\varphi_{inl}^{(\tt m)};\varphi_{jnl}^{(\tt s)}$ & Path loss from $U_{nl}$ to MBS $i$ and SCB $j$ \\
$\mathcal{A}_{\tt m}^{\tt L}; \mathcal{A}_{\tt m}^{\tt L}; \mathcal{A}_{\tt s}^{\tt L}; \mathcal{A}_{\tt s}^{\tt NL}$ & Probability that the typical user is associated with the MBS in LoS and NLoS, and with SCB in LoS and NLoS\\
$R_{\tt m}^{\tt L}; R_{\tt m}^{\tt NL}; R_{\tt s}^{\tt L}; R_{\tt s}^{\tt NL}$ & Distance between a user and its serving BS when this user is associated with the MBS in LoS and NLoS, and with SCB in LoS and NLoS \\
$\mathcal{U}_i^{\tt m}; \mathcal{U}_j^{\tt s}$ & Collection of users associated with MBS $i$ and SCB $j$\\
$\tau; p_p; \sigma^2$ & Length of uplink pilots; pilot transmit power; noise variance\\
\hline
\end{tabular}
\end{center}\vspace{-0.63cm}
\end{table}%

\subsection{Network Topology}

\begin{figure}[!t]
  \centering
  \includegraphics[scale=0.6]{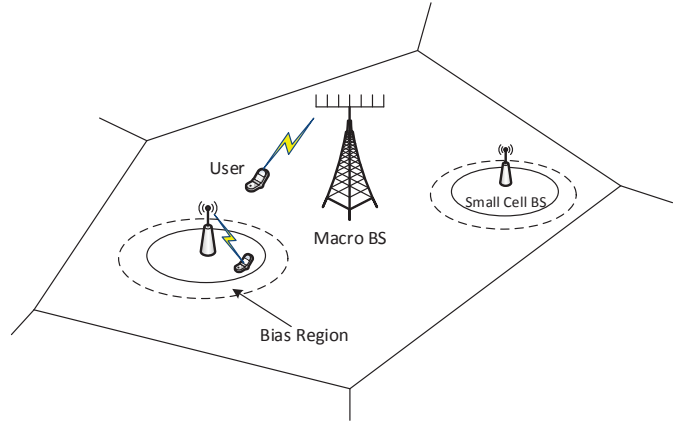}
  \caption{A two-tier heterogenous network utilizing a mix of macro and small cell BSs.}\label{fig 1}
\end{figure}

We consider the downlink of a two-tier heterogenous cellular network, where high-power macro BSs (MBSs) are overlaid with successively denser and lower-power small cell BSs (SCBs), as illustrated in Fig. \ref{fig 1}. We assume that the MBSs and SCBs are deployed on a plane according to independent PPPs $\Phi_{\tt m}$ and $\Phi_{\tt s}$ with spatial densities $\lambda_{\tt m}$ and $\lambda_{\tt s}$, respectively. All MBSs and SCBs are equipped with $M_{\tt m}$ and $M_{\tt s}$ antennas, respectively, whereas each MBS transmits with power $P_{\tt m}$, and each SCB has transmit power to be $P_{\tt s}$. In its light of high spectral utilization, we consider a co-channel deployment of small cells with the macro cell tier, i.e., MBSs and SCBs share the same frequency band for transmission. We model the mobile users as another independent PPP $\Phi_{\tt u}$ with spatial density $\lambda_{\tt u}$. Additionally,
we assume $\lambda_{\tt u}$ is much larger than $\lambda_{\tt m}$, and each MBS has at least $N$ users in its coverage for analytical tractability\footnote{Note that removing this assumption does not change the main outcomes of this paper since the probability of having less number of users than $N$ is very small for large $\lambda_{\tt u}$ \cite{dhillon13,yang15,bai15}.}.
In each time-frequency resource block, a MBS first schedule $N$ users in its coverage based on the average received power and then these users are flexibly associated with the MBS or SCB according to the association policy described in Section \ref{subsec: user association policy}. The user $n$ in macro cell $l$ is denoted as $U_{nl}$.

\subsection{Propagation Environment}
We model the channels between any pair of antennas as independent and identically distributed (i.i.d.) and quasi-static, i.e., the channel is constant during a sufficiently
long coherence block, and varies independently from block to block. Moreover, we assume that each channel is narrowband and affected by two attenuation components, namely small-scale Rayleigh fading, and large-scale path loss. Regarding the practical LoS and NLoS transmissions, we model the path loss with two parts, i.e., the LoS and NLoS path. More formally, the path loss $\varphi_{inl}$ between $U_{nl}$ and BS $i$ can be written as follows
\begin{equation}\label{beta model}
\varphi_{inl} =\left\{
\begin{array}{ll}
\varphi_{inl}^{ \tt L}=L^{\tt L} r_{inl}^{ -\alpha^{\tt L}}, & \text{if LoS},\\
\varphi_{inl}^{ \tt NL}=L^{\tt NL} r_{inl}^{ -\alpha^{\tt NL}}, & \text{if NLoS},\\
\end{array}\right.
\end{equation}
where $r_{inl}$ is the distances between user $U_{nl}$ and BS $i$, $L^{\tt L}$ and $L^{\tt NL}$ denote the path losses evaluated at a reference distance $1$ for LoS and NLoS, respectively, and $\alpha^{\tt L}$ and $\alpha^{\tt NL}$ are the LoS and NLoS path loss exponents, respectively.
As the probability of a wireless link being LoS or NLoS is mainly affected by the distance between the transmitter and receiver, we model such probability as a homogeneous event in the following analysis \cite{bai14}. Therefore, the channel coefficient between $U_{nl}$ and the $m$-th antenna of MBS $i$ can be formulated as
\begin{equation}\label{g los nlos}
g_{minl}^{(\tt m)}=\left\{
\begin{array}{ll}
h_{minl}^{(\tt m)} \sqrt{\varphi_{inl}^{(\tt m),\tt L}}, & \text{with probability } \bbP^{\tt L}_{\tt m}(r^{(\tt m)}_{inl}),\\
h_{minl}^{(\tt m)} \sqrt{\varphi_{inl}^{(\tt m),\tt NL}}, & \text{with probability } 1-\bbP^{\tt L}_{\tt m}(r^{(\tt m)}_{inl}),\\
\end{array}\right.
\end{equation}
and the channel coefficient between $U_{nl}$ and the $m$-th antenna of SCB $j$ can be formulated as
\begin{equation}\label{g j los nlos}
g_{mjnl}^{(\tt s)}=\left\{
\begin{array}{ll}
h_{mjnl}^{(\tt s)} \sqrt{\varphi_{jnl}^{(\tt s),\tt L}}, & \text{with probability } \bbP^{\tt L}_{\tt s}(r^{(\tt s)}_{jnl}),\\
h_{mjnl}^{(\tt s)} \sqrt{\varphi_{jnl}^{(\tt s),\tt NL}}, & \text{with probability } 1-\bbP^{\tt L}_{\tt s}(r^{(\tt s)}_{jnl}),\\
\end{array}\right.
\end{equation}
where $h_{minl}^{(\tt m)}$ and $h_{mjnl}^{(\tt s)}$ denote the small-scale fading while $h_{minl}^{(\tt m)}$, $h_{mjnl}^{(\tt s)} \sim \mathcal{CN}(0,1)$, and $\bbP^{\tt L}_{\tt m}(\cdot)$ and $\bbP^{\tt L}_{\tt s}(\cdot)$ are the LoS probabilities function with the MBS and SCB, respectively. Note that the LoS probabilities can be different for macro and small cells due to the assorted propagation environment as well as various antenna heights.

\subsection{User Association Policy}\label{subsec: user association policy}
From the perspective of load balancing, we adopt cell range expansion for user associations in this network. Specifically, all the SCBs employ a bias factor $B$ for the control of cell range expansion, and users associate to the BS that provides the largest average biased received power. Note that with the existence of LoS path, it is possible that a user is associated to a far away LoS BS instead of a nearby NLoS BS. Due to the stationary property of PPP, we can evaluate the performance of a typical user located at the origin, denoted as $U_{00}$, thanks to Slivyark's theorem. As such, the average biased-received power of the typical user from the MBS $i$ is
\begin{equation}\label{brp macro}
\mathcal{P}_{\tt m}(r^{(\tt m)}_{i00})=\left\{
\begin{array}{ll}
P_{\tt m} L^{\tt L} r_{i00}^{(\tt m)-\alpha^{\tt L}} , & \text{with probability } \bbP^{\tt L}_{\tt m}(r^{(\tt m)}_{i00}),\\
 P_{\tt m} L^{\tt NL} r_{i00}^{(\tt m)-\alpha^{\tt NL}}, & \text{with probability } 1-\bbP^{\tt L}_{\tt m}(r^{(\tt m)}_{i00}),\\
 \end{array}\right.
\end{equation}
and the average biased-received power of the typical user from the SCB $j$ is
\begin{equation}\label{brp scb}
\mathcal{P}_{\tt s}(r^{(\tt s)}_{j00})=\left\{
\begin{array}{ll}
 P_{\tt s} L^{\tt L} r_{j00}^{(\tt s)-\alpha^{\tt L}} B, & \text{with probability } \bbP^{\tt L}_{\tt s}(r_{j00}^{(\tt s)}),\\
P_{\tt s} L^{\tt NL} r_{j00}^{(\tt s)-\alpha^{\tt NL}} B, & \text{with probability } 1-\bbP^{\tt L}_{\tt s}(r_{j00}^{(\tt s)}),\\
 \end{array}\right.
\end{equation}
where $B$ is the small-cell bias factor, which can be set greater or smaller than one to extend or shrink the coverage.

\section{Achievable Downlink Rate}\label{sec: achievable downlink rate}
In this section, we analyze the achievable downlink rate of the HCN with LoS/NLoS transmissions. Particularly, we utilize tools from stochastic geometry to derive the user association probabilities, the distribution of serving distance, and finally the tight approximation of achievable downlink rate. The resulting analysis captures the effects of LoS/NLoS transmissions, pilot contamination, and random network topology. For a better readability, most proofs and mathematical derivations have been relegated to the Appendix.

\subsection{Association Probabilities}
With the presence of LoS/NLoS transmission, coverage area of BSs no longer form weighted Voronoi cells because a user can associate to a far away BS with LoS path instead of a nearby BS with NLoS path. To analyze this more challenging user association, we start with a decomposition of the PPPs $\Phi_{\tt m}$ and $\Phi_{\tt s}$.
More precisely, if an MBS has a LoS path to the typical user located at the origin, we classify it as in the LoS MBS set $\Phi_{\tt m}^{\tt L}$, otherwise, we put it into the set of NLoS MBS $\Phi_{\tt m}^{\tt NL}$. Given that this operation is performed independently for each point in $\Phi_{\tt m}$, from the Thinning Theorem \cite[Theorem 2.36]{haenggi12}, it follows that $\Phi_{\tt m}^{\tt L}$ and $\Phi_{\tt m}^{\tt NL}$ are two independent inhomogeneous PPPs with densities $\lambda_{\tt m}\bbP_{\tt m}^{\tt L}(r)$ and $\lambda_{\tt m}(1-\bbP_{\tt m}^{\tt L}(r))$, respectively, where $r$ stands for the distance from an MBS to the typical user. Similarly, we can also decompose $\Phi_{\tt s}$ into the sets of LoS SCB $\Phi_{\tt s}^{\tt L}$ and NLoS SCB $\Phi_{\tt s}^{\tt NL}$, which are two inhomogeneous PPPs with densities $\lambda_{\tt s}\bbP_{\tt s}^{\tt L}(r)$ and $\lambda_{\tt s}(1-\bbP_{\tt s}^{\tt L}(r))$, respectively.


The distance from the typical user to its nearest BS in each $\Phi_{\tt m}^{\tt L}$, $\Phi_{\tt m}^{\tt NL}$, $\Phi_{\tt s}^{\tt L}$, and $\Phi_{\tt s}^{\tt NL}$ are denoted as $S_{\tt m}^{\tt L}$, $S_{\tt m}^{\tt NL}$, $S_{\tt s}^{{\tt L}}$, and $S_{\tt s}^{{\tt NL}}$, respectively.
We further define the event that the typical user is associated with the MBS in LoS and NLoS, and associated with the SCB in LoS and NLoS, as $E_{\tt m}^{\tt L}$, $E_{\tt m}^{\tt NL}$, $E_{\tt s}^{\tt L}$, and $E_{\tt s}^{\tt NL}$, respectively. The following theorem provides the probabilities of these events.

\begin{theorem}\label{theorem 1}
Probabilities that the typical user is associated with the MBS in LoS and NLoS path are given by
\begin{align}
\mathcal{A}_{\tt m}^{\tt L} \triangleq
\bbP\left[E_{\tt m}^{\tt L}\right]& =2 \pi \lambda_{\tt m} \int_0^\infty r \bbP_{\tt m}^{\tt L}(r) \zeta_1\left(r,k_1 r^\frac{\alpha^{\tt L}}{\alpha^{\tt NL}}\right)\zeta_2\left(k_2 r, k_1 k_3 r^\frac{\alpha^{\tt L}}{\alpha^{\tt NL}}\right) dr, \label{Prob EmL} \\
\mathcal{A}_{\tt m}^{\tt NL} \triangleq
\bbP\left[E_{\tt m}^{\tt NL}\right]& =2 \pi \lambda_{\tt m} \int_0^\infty r (1-\bbP_{\tt m}^{\tt L}(r)) \zeta_1\left(k_4 r^\frac{\alpha^{\tt NL}}{\alpha^{\tt L}},r\right)\zeta_2\left(k_2 k_4 r^\frac{\alpha^{\tt NL}}{\alpha^{\tt L}}, k_3 r\right) dr, \label{Prob EmNL}
\end{align}
and probabilities that the typical user is associated with the SCB in LoS and NLoS path are given by
\begin{align}
\mathcal{A}_{\tt s}^{\tt L} \triangleq
\bbP\left[E_{\tt s}^{\tt L}\right]&=2 \pi \lambda_{\tt s} \int_0^\infty r \bbP_{\tt s}^{\tt L}(r) \zeta_1\left(r/k_2,k_1/k_3 r^\frac{\alpha^{\tt L}}{\alpha^{\tt NL}}\right)\zeta_2\left(r, k_1 r^\frac{\alpha^{\tt L}}{\alpha^{\tt NL}}\right) dr, \label{Prob EsL}\\
\mathcal{A}_{\tt s}^{\tt NL} \triangleq
\bbP\left[E_{\tt s}^{\tt NL}\right]&=2 \pi \lambda_{\tt s} \int_0^\infty r (1-\bbP_{\tt s}^{\tt L}(r)) \zeta_1\left(k_4/k_2 r^\frac{\alpha^{\tt NL}}{\alpha^{\tt L}},r/k_3\right)\zeta_2\left( k_4 r^\frac{\alpha^{\tt NL}}{\alpha^{\tt L}},  r\right) dr, \label{Prob EsNL}
\end{align}
where
\begin{align}
\zeta_1(x_1,x_2) &\triangleq \exp\left(-\int_0^{x_1} \lambda_{\tt m} \bbP_{\tt m}^{\tt L}(u)2 \pi u du  - \int_0^{x_2} \lambda_{\tt m} (1-\bbP_{\tt m}^{\tt L}(u)) 2 \pi u du \right),\\
\zeta_2(x_1,x_2) &\triangleq \exp\left(-\int_0^{x_1} \lambda_{\tt s} \bbP_{\tt s}^{\tt L}(u)2 \pi u du  - \int_0^{x_2} \lambda_{\tt s} (1-\bbP_{\tt s}^{\tt L}(u)) 2 \pi u du \right),
\end{align}
with $k_1 \triangleq\left({L^{\tt NL}}/{L^{\tt L}}\right)^{1/\alpha^{\tt NL}}$, $k_2 \triangleq \left({B P_{\tt s}}/{P_{\tt m}}\right)^{1/\alpha^{\tt L}}$, $k_3 \triangleq \left({B P_{\tt s}}/{P_{\tt m}}\right)^{1/\alpha^{\tt NL}}$, and $k_4 \triangleq \left({L^{\tt L}}/{L^{\tt NL}}\right)^{1/\alpha^{\tt L}}$.
\end{theorem}
\begin{IEEEproof}
See Appendix \ref{proof of theorem 1}.
\end{IEEEproof}


The accuracy of Theorem \ref{theorem 1} will be validated in Fig. \ref{fig compare_assoProb}. From Theorem \ref{theorem 1}, it is easy to obtain the probability that the typical user is associated with an MBS as
$\mathcal{A}_{\tt m}=\mathcal{A}_{\tt m}^{\tt L}+\mathcal{A}_{\tt m}^{\tt NL}$,
and with an SCB as
$\mathcal{A}_{\tt s}=\mathcal{A}_{\tt s}^{\tt L}+\mathcal{A}_{\tt s}^{\tt NL}$.
 Based on the above results, we can further derive the average number of users associated with each BS as follows.

\begin{corollary}\label{corollary 1}
The average number of users associated with an MBS is $\mathcal{N}_{\tt m} = \mathcal{A}_{\tt m} N$, and with an SCB is $\mathcal{N}_{\tt s} = \mathcal{A}_{\tt s} \lambda_{\tt m} N /\lambda_{\tt s}$.
\end{corollary}
\begin{IEEEproof}
See Appendix \ref{proof of corollary 1}.
\end{IEEEproof}

\begin{remark}\label{remark 1}
 Larger $\lambda_{\tt s}$ results in higher probability for a user to be associated with SCB and reduces the probability associated with MBS, i.e., $\mathcal{A}_{\tt s}$ grows and $\mathcal{A}_{\tt s}$ decreases. Hence, as $\lambda_{\tt s}$ increases, the decline of $\mathcal{N}_{\tt m}$ is obvious from Corollary \ref{corollary 1}, while the behavior of $\mathcal{N}_{\tt s}$ is hard to fathom. In Section \ref{sec: numerical results}, we observe that $\mathcal{A}_{\tt s}$ is almost a linear function of $\lambda_{\tt s}$ with no offset, i.e., $\mathcal{A}_{\tt s} \approx a \lambda_{\tt s}$, where $a$ is a positive value. Following Corollary \ref{corollary 1}, we get that $\mathcal{N}_{\tt s} \approx a \lambda_{\tt m} N$, which means the average number of users associated with each SCB remains unchanged as $\lambda_{\tt s}$ increases.
\end{remark}

 When the typical user is associated with the MBS in a LoS or NLoS path, the distance between the user and its serving BS are denoted as $R_{\tt m}^{\tt L}$ and $R_{\tt m}^{\tt NL}$, respectively. Similarly, we use $R_{\tt s}^{\tt L}$ and $R_{\tt s}^{\tt NL}$ to represent the distance between the typical user and its serving SCB, when the association is through LoS or NLoS path, respectively. The next theorem provides the probability density function (pdf) for each of these distances.

\begin{theorem}\label{theorem 2}
The pdf of $R_{\tt m}^{\tt L}$, $R_{\tt m}^{\tt NL}$, $R_{\tt s}^{\tt L}$ and $R_{\tt s}^{\tt NL}$ are given as follows:
\begin{align}
f_{R_{\tt m}^{\tt L}} (r) &= \frac{2 \pi \lambda_{\tt m}}{\mathcal{A}_{\tt m}^{\tt L}} r \bbP_{\tt m}^{\tt L}(r) \zeta_1\left(r,k_1 r^\frac{\alpha^{\tt L}}{\alpha^{\tt NL}}\right)\zeta_2\left(k_2 r, k_1 k_3 r^\frac{\alpha^{\tt L}}{\alpha^{\tt NL}}\right),\label{f mL}\\
f_{R_{\tt m}^{\tt NL}} (r) &=\frac{ 2 \pi \lambda_{\tt m}}{\mathcal{A}_{\tt m}^{\tt NL}}  r \left(1-\bbP_{\tt m}^{\tt L}(r)\right) \zeta_1\left(k_4 r^\frac{\alpha^{\tt NL}}{\alpha^{\tt L}},r\right)\zeta_2\left(k_2 k_4 r^\frac{\alpha^{\tt NL}}{\alpha^{\tt L}}, k_3 r\right) dr,\label{f mNL}\\
f_{R_{\tt s}^{\tt L}} (r) &= \frac{2 \pi \lambda_{\tt s}}{\mathcal{A}_{\tt s}^{\tt L}}  r \bbP_{\tt s}^{\tt L}(r) \zeta_1\left(r/k_2,k_1/k_3 r^\frac{\alpha^{\tt L}}{\alpha^{\tt NL}}\right)\zeta_2\left(r, k_1 r^\frac{\alpha^{\tt L}}{\alpha^{\tt NL}}\right),\label{f sL}\\
f_{R_{\tt s}^{\tt NL}} (r) &=\frac{2 \pi \lambda_{\tt s}}{\mathcal{A}_{\tt s}^{\tt NL}} r \left(1-\bbP_{\tt s}^{\tt L}(r)\right) \zeta_1\left(k_4/k_2 r^\frac{\alpha^{\tt NL}}{\alpha^{\tt L}},r/k_3\right)\zeta_2\left( k_4 r^\frac{\alpha^{\tt NL}}{\alpha^{\tt L}},  r\right).\label{f sNL}
\end{align}
\end{theorem}
\begin{IEEEproof}
See Appendix \ref{proof of theorem 2}.
\end{IEEEproof}

The above results will be applied in the derivation of achievable downlink rate. In the next section, we investigate the channel estimation procedure from uplink training.

\subsection{Uplink Training}
During a dedicated uplink training phase, users in each macro cell simultaneously transmit mutually orthogonal pilot sequences which allow the BSs to estimate channels of users associated with them. Due to the limited pilot length, we further assume that the same set of orthogonal pilot sequences is reused in every macro cell.
In particular, the MBS assigns orthogonal pilots of length $\tau$ symbols for the $N$ users in its cell ($\tau \ge N$), and notifies each SCB the pilot sequence of users associated with it.

 The pilot sequence used by $U_{nl}$ is expressed as a $\tau \times 1$ vector $\sqrt {\tau} \bs{\epsilon}_{nl}$, which satisfies $\bs{\epsilon}_{nl}^H \bs{\epsilon}_{cl}=\delta[n-c]$, with $\delta[\cdot]$ being the Kronecker delta function.
Furthermore, we assume that for any $i \ne l$, $\bs{\epsilon}_{ni}=\bs{\epsilon}_{nl}$, i.e, the $n$-th user in every macro cell has the same pilot sequence. By transmitting these pilot signals over $\tau$ symbols in the uplink, the collective received pilot signal at MBS $i$ can be expressed as
\begin{equation}\label{received pilot macro}
\mathbf{Y}_i^{(\tt m)}= \sqrt{\tau p_p}\sum\limits_{l\in \Phi_{\tt m}}\sum\limits_{n=1}^N\bg^{(\tt m)}_{inl} \bs{\epsilon}_{nl}+\mathbf{N}_i^{(\tt m)},
\end{equation}
where $p_p$ denotes the pilot power, $\bg_{inl}^{(\tt m)}=[g^{(\tt m)}_{1inl},\ldots,g^{(\tt m)}_{M_{\tt m}inl}]^T$ is the channel vector from $U_{nl}$ to the MBS $i$, and $\mathbf{N}_i^{(\tt m)}$ represents the $M_{\tt m} \times \tau$ additive white Gaussian noise (AWGN) matrix with i.i.d. zero-mean elements and variance $\sigma^2$. Similarly, the $M_{\tt s}\times \tau$ noisy pilot matrix at SCB $j$ can be written as
\begin{equation}\label{received pilot scb}
\mathbf{Y}_j^{(\tt s)}= \sqrt{\tau p_p}\sum\limits_{l\in \Phi_{\tt m} } \sum\limits_{n=1}^N \bg^{(\tt s)}_{jnl}\bs{\epsilon}_{nl}+\mathbf{N}^{(\tt s)}_j,
\end{equation}
where $\bg^{(\tt s)}_{jnl}=[g^{(\tt s)}_{1jnl},\ldots,g^{(\tt s)}_{M_{\tt s}jnl}]^T$, and $\mathbf{N}^{(\tt s)}_j$ is an AWGN matrix.

Each BS estimates a user channel through multiplying the received pilot signal by the corresponding pilot sequence used by this user. In this work, we adopt the minimum mean-square-error (MMSE) estimation method. As a result, the estimated channel vector $\bhg^{(\tt m)}_{ini}$ is given by
  \begin{equation}\label{gini estimate}
 \bhg^{(\tt m)}_{ini}=\eta^{(\tt m)}_{ini} \frac{1}{\sqrt{\tau p_p}}\mathbf{Y}^{(\tt m)}_i \bs \epsilon_{ni}^H,
  \end{equation}
where $\eta^{(\tt m)}_{ini}\triangleq {\varphi^{(\tt m)}_{ini}}/\left({\sum\limits_{l\in \Phi_{\tt m}}\varphi^{(\tt m)}_{inl}+\frac{\sigma^2}{\tau p_p}}\right)$.
  Let $\btg^{(\tt m)}_{ini}=\bhg^{(\tt m)}_{ini}-\bg^{(\tt m)}_{ini}$ denote the channel estimation error. Then, we know that the elements of $\btg^{(\tt m)}_{ini}$ follows $\mathcal{CN}(0,\varphi^{(\tt m)}_{ini}(1-\eta^{(\tt m)}_{ini}))$.
Similarly, the MMSE estimate for channel vector $\bg_{jni}^{(\tt s)}$ is given by
  \begin{equation}\label{gjni estimate}
  \bhg^{(\tt s)}_{jni}= \eta_{jni}^{(\tt s)}\frac{1}{\sqrt{\tau p_p}}\mathbf{Y}^{(\tt s)}_j \bs \epsilon_{ni}^H,
  \end{equation}
  where $\eta_{jni}^{(\tt s)} \triangleq {\varphi^{(\tt s)}_{jni}}/\left({\sum\limits_{l\in \Phi_{\tt m}}\varphi^{(\tt s)}_{jnl}+\frac{\sigma^2}{\tau p_p}}\right)$, and elements of channel estimation error $\btg^{(\tt s)}_{ini}=\bhg^{(\tt s)}_{ini}-\bg^{(\tt s)}_{ini}$ follows distribution $\mathcal{CN}(0,\varphi^{(\tt s)}_{jni}(1-\eta^{(\tt s)}_{jni}))$.

Note that due to pilot reuse, the estimated channel vector is polluted by channels from users in other cells who share the same pilot, thus causing the pilot contamination. As mentioned in \cite{marzetta10}, pilot contamination is a main limiting factor for the performance of massive MIMO, and the result of this impact with LoS/NLoS transmissions will be further explored in Section \ref{sec: numerical results}.

\subsection{Downlink Data Transmission}
Let
$\mathcal{U}_{i}^{\tt m}$ and $\mathcal{U}_{j}^{\tt s}$ be the collection of users associated with the MBS $i$ and the SCB $j$, respectively, and
$\left|\mathcal{U}_{i}^{\tt m}\right|$ and $\left|\mathcal{U}_{j}^{\tt s}\right|$ denote the corresponding cardinalities.
 Each BS uses the estimated channel obtained from uplink training to establish the downlink precoding vector. Then, the received signal at the typical user can be written as
 \begin{equation}\label{received signal downlink}
  s_0=\sqrt{P_{\tt m}} \sum\limits_{l \in \Phi_{\tt m}}\sum\limits_{U_{nl}\in \mathcal{U}_{l}^{\tt m}}\bg^{{(\tt m)}H}_{l00}\mbf^{(\tt m)}_{lnl} x^{(\tt m)}_{lnl}+ \sqrt{P_{\tt s}}\sum\limits_{j \in \Phi_{\tt s}} \sum\limits_{l \in \Phi_{\tt m}} \sum\limits_{U_{nl}\in \mathcal{U}^{\tt s}_{j}}\bg^{{(\tt s)}H}_{j00}\mbf^{(\tt s)}_{jnl} x^{(\tt s)}_{jnl}+n_0,
  \end{equation}
where $\mbf^{(\tt m)}_{lnl}$ is the $M_{\tt m} \times 1$ precoding vector of MBS $l$ to $U_{nl}$, and $\mbf^{(\tt s)}_{jnl}$ is the $M_{\tt s} \times 1$ precoding vector of SCB $j$ to $U_{nl}$, $x^{(\tt m)}_{lnl}$ and $x^{(\tt m)}_{jnl}$ are the signals intended for $U_{nl}$ from MBS $l$ and the SCB $j$, respectively, and $n_0$ is the AWGN.

We consider the maximal-ratio-transmission precoding in this work, and the precoding vector $\mbf_{lnl}^{(\tt m)}$ is given by
\begin{equation}\label{flnl m}
\mbf_{lnl}^{(\tt m)} = \kappa_l^{(\tt m)} {\bhg_{lnl}^{(\tt m)}}/{\sqrt{\bbE_{\bh}\left\{\left\|\bhg_{lnl}^{(\tt m)}\right\|^2\right\}}},
\end{equation}
where $\bbE_{\bh}$ means the average over fast fading, and $\kappa^{(\tt m)}_{l}$ is a power normalization factor which conforms to the following constraint\footnote{Here, we consider an average power normalization over the fast fading \cite{yang13}, which can be an instantaneous constraint for each large-scale realization.}
\begin{equation}\label{Fll constraint}
\bbE_{\bh} \left\{{\tt tr}\left(\mathbf{F}^{(\tt m)}_{l} \mathbf{F}^{{(\tt m)}H}_{l} \right)\right\}=1,
\end{equation}
where $\mathbf{F}^{(\tt m)}_{l}=\left\{\left[\ldots,\mbf^{(\tt m)}_{lnl},\ldots\right]~| U_{nl} \in \mathcal{U}^{\tt m}_{l}\right\}$. Therefore, we get that
$\kappa_l^{(\tt m)} = \sqrt{1/\left|\mathcal{U}_l^{\tt m}\right|}$.
Similarly, the precoding vector $\mbf_{jnl}^{(\tt s)}$ for SCB is given as
\begin{equation}\label{fjnl s}
\mbf_{jnl}^{(\tt s)} = \kappa_j^{(\tt s)} {\bhg_{jnl}^{(\tt s)}}/{\sqrt{\bbE_{\bh}\left\{\left\|\bhg_{jnl}^{(\tt s)}\right\|^2\right\}}},
\end{equation}
where $\kappa_j^{(\tt s)} = \sqrt{1/\left|\mathcal{U}_j^{\tt s}\right|}$.

To this end, if the typical user is associated with MBS $0$,
 the downlink SINR is given by
   \begin{align}\label{sinr macro}
      {\tt SINR}_{\tt m}=
      \frac{P_{\tt m} \left|\bhg_{000}^{{(\tt m)}H}\mbf^{(\tt m)}_{000}\right|^2}{{\tt I}_{\tt m}+ \sigma^2 },
      \end{align}
  where
  \begin{multline}\label{I m}
  {\tt I}_{\tt m} \triangleq P_{\tt m} \nsp \nsp \nsp  \nsp \sum\limits_{U_{n0}\in \mathcal{U}^{\tt m}_{0} \backslash U_{00}} \nsp    \left|\bhg_{000}^{{(\tt m)}H} \mbf^{(\tt m)}_{0n0} \right|^2+ P_{\tt m} \nsp \nsp \sum\limits_{U_{n0}\in \mathcal{U}^{\tt m}_{0}}  \left|\btg_{000}^{{(\tt m)}H} \mbf^{(\tt m)}_{0n0} \right|^2 \\
  + P_{\tt m} \nsp \nsp \sum\limits_{l \in \Phi_{\tt m}\backslash 0}\sum\limits_{U_{nl}\in \mathcal{U}^{\tt m}_{l}} \left|\bg_{l00}^{{(\tt m)}H}\mbf^{(\tt m)}_{lnl}\right|^2 +  P_{\tt s}\sum\limits_{j \in \Phi_{\tt s}}\sum\limits_{l \in \Phi_{\tt m}} \sum\limits_{U_{nl}\in \mathcal{U}^{\tt s}_j} \left|\bg_{j00}^{{(\tt s)}H}\mbf^{(\tt s)}_{jnl}\right|^2.
  \end{multline}
   If the typical user is associated with the SCB, denoted as $q_0$,
the downlink SINR is given by
      \begin{align}\label{sinr scb}
      {\tt SINR}_{\tt s}= \frac{P_{\tt s} \left|\bhg^{{(\tt s)}H}_{q_000} \mbf^{(\tt s)}_{q_000}\right|^2}{{\tt I}_{\tt s}+  \sigma^2},
      \end{align}
      where
      \begin{multline}\label{I s}
      {\tt I}_{\tt s} \triangleq P_{\tt s} \nsp \sum\limits_{l \in \Phi_{\tt m}}\left(\sum\limits_{U_{nl}\in \mathcal{U}_{q_0}^{\tt s}\backslash U_{00} }\nsp \nsp \nsp \nsp \left|\bhg_{q_000}^{^{(\tt s)}H} \mbf^{(\tt s)}_{q_0nl} \right|^2 +  \nsp \nsp \nsp \sum\limits_{U_{nl}\in \mathcal{U}_{q_0}^{\tt s}}\nsp \nsp  \left|\btg_{q_000}^{^{(\tt s)}H} \mbf^{(\tt s)}_{q_0nl} \right|^2 \right) \\ + P_{\tt m} \nsp \sum\limits_{l \in \Phi_{\tt m}}\sum\limits_{U_{nl}\in \mathcal{U}^{\tt m}_{l}} \left|\bg_{l00}^{{(\tt m)}H}\mbf^{(\tt m)}_{lnl}\right|^2 +  P_{\tt s}\nsp \nsp \nsp \sum\limits_{j \in \Phi_{\tt s}\backslash q_0}  \sum\limits_{l \in \Phi_{\tt m}} \sum\limits_{U_{nl}\in \mathcal{U}_{j}^{\tt s}} \left|\bg_{j00}^{{(\tt s)}H}\mbf^{(\tt s)}_{jnl}\right|^2.
      \end{multline}

\subsection{Approximation of Achievable Downlink Rate}\label{sec: closed-form achievable downlink rate}

Based on the SINR given in \eqref{sinr macro} and \eqref{sinr scb}, the achievable downlink rate of the typical user can be expressed as
\begin{align}\label{rate comp}
\bbR = \bbE\left\{\log_2\left(1+{\tt SINR}\right)\right\}&=\bbE\left\{\log_2\left(1+{{\tt SINR}_{\tt m}^{\tt L}}\right)\right\}\mathcal{A}_{\tt m}^{\tt L}+\bbE\left\{\log_2\left(1+{{\tt SINR}_{\tt m}^{\tt NL}}\right)\right\}\mathcal{A}_{\tt m}^{\tt NL}\notag\\
&+\bbE\left\{\log_2\left(1+{{\tt SINR}_{\tt s}^{\tt L}}\right)\right\}\mathcal{A}_{\tt s}^{\tt L}+\bbE\left\{\log_2\left(1+{{\tt SINR}_{\tt s}^{\tt NL}}\right)\right\}\mathcal{A}_{\tt s}^{\tt NL},
\end{align}
where ${\tt SINR}_{\tt m}^{\tt L}$, ${\tt SINR}_{\tt m}^{\tt NL}$, ${\tt SINR}_{\tt s}^{\tt L}$, and ${\tt SINR}_{\tt s}^{\tt NL}$ denote the received SINR of the typical user when it is associated with the MBS in LoS path, NLoS path, and with the SCB in LoS path and NLoS path, respectively. To facilitate the rate derivation, we make the following assumptions.
\begin{assumption}\label{assumption 1}
We approximate the coverage region of the MBS $\mathcal{M}$ as a ball centered at $\mathcal{M}$ with radius $C_{\tt v} = 1/\sqrt{\pi \lambda_{\tt m}}$ {\normalfont \cite{singh13}}.
\end{assumption}

Let $\mathcal{N}_{n}$ be the point process formed by locations of the user $n$ in each macro cell. Note that $\mathcal{N}_n$ is a perturbation of the process $\Phi_{\tt m}$ and thus not a PPP. Obtaining the exact correlation between $\Phi_{\tt m}$ and $\mathcal{N}_{n}$ requires complicated mathematical derivations and is highly intractable. Therefore, we use the similar method in \cite{singh15} to model the interfering users in $\mathcal{N}_{n}$, denoted by $\mathcal{N}'_{n}$, as an inhomogeneous PPP. Motivated by the fact that the probability that a user scheduled by the MBS $\mathcal{M}$ in a LoS path is
$\zeta_1\left(r,k_1 r^\frac{\alpha^{\tt L}}{\alpha^{\tt NL}}\right)$,
and in a NLoS path is
$\zeta_1\left(k_4 r^\frac{\alpha^{\tt NL}}{\alpha^{\tt L}},r\right)$,
where $r$ is the distance of this user to $\mathcal{M}$, we make the following assumption.

\begin{assumption}\label{assumption 2}
The point process $\mathcal{N}'_n$ can be approximated as an inhomogeneous PPP with density being
\begin{equation}\label{intensity N0}
\lambda_{\mathcal{N}'_n}(r) = \lambda_{\tt m} \left[1-\zeta_1\left(r,k_1 r^\frac{\alpha^{\tt L}}{\alpha^{\tt NL}}\right)-\zeta_1\left(k_4 r^\frac{\alpha^{\tt NL}}{\alpha^{\tt L}},r\right)\right].
\end{equation}
\end{assumption}
Moreover, for $n_1 \ne n_2$, $\mathcal{N}'_{n_1}$ and $\mathcal{N}'_{n_2}$ are independent.

Based on all the analysis and assumptions mentioned above, a tight approximation of the achievable downlink rate is given in the following theorem.
\begin{theorem}\label{theorem 3}
The achievable downlink rate of the typical user is approximated by
\begin{align}\label{rate approximation}
\bbR \approx \tilde \bbR &= \mathcal{A}_{\tt m}^{\tt L} \int_0^\infty \nsp \int_0^\infty \frac{e^{-z}}{z\ln 2}  \Xi\left(r,k_1 r^\frac{\alpha^{\tt L}}{\alpha^{\tt NL}},k_2r, k_1 k_3 r^\frac{\alpha^{\tt L}}{\alpha^{\tt NL}}\right) \Psi\left(P_{\tt m},\mathbb{N}_{\tt m}, M_{\tt m},\chi_1,\frac{L^{\tt L}}{ r^{\alpha^{\tt L}}}\right)f_{R_{\tt m}^{\tt L}}(r) dz dr \notag\\
&+\mathcal{A}_{\tt m}^{\tt NL} \int_0^\infty \nsp \int_0^\infty \frac{e^{-z}}{z\ln 2} \Xi\left(k_4 r^\frac{\alpha^{\tt NL}}{\alpha^{\tt L}}, r, k_2 k_4r^\frac{\alpha^{\tt NL}}{\alpha^{\tt L}}, k_3 r \right) \Psi\left(P_{\tt m},\mathbb{N}_{\tt m}, M_{\tt m},\chi_1,\frac{L^{\tt NL}}{ r^{\alpha^{\tt NL}}}\right)f_{R_{\tt m}^{\tt NL}}(r) dz dr\notag\\
&+\mathcal{A}_{\tt s}^{\tt L} \int_0^\infty \nsp \int_0^\infty \frac{e^{-z}}{z\ln 2}\Xi\left(\frac{r}{k_2},\frac{k_1}{k_3} r^\frac{\alpha^{\tt L}}{\alpha^{\tt NL}},r, k_1 r^\frac{\alpha^{\tt L}}{\alpha^{\tt NL}}\right) \Psi\left(P_{\tt s},\mathbb{N}_{\tt s}, M_{\tt s},\chi_2,\frac{L^{\tt L}}{ r^{\alpha^{\tt L}}}\right)f_{R_{\tt s}^{\tt L}}(r) dz dr \notag\\
&+\mathcal{A}_{\tt s}^{\tt NL} \int_0^\infty \nsp \int_0^\infty \frac{e^{-z}}{z\ln 2}  \Xi\left(\frac{k_4}{k_2} r^\frac{\alpha^{\tt NL}}{\alpha^{\tt L}}, \frac{r}{k_3}, k_4r^\frac{\alpha^{\tt NL}}{\alpha^{\tt L}}, r \right) \Psi\left(P_{\tt s},\mathbb{N}_{\tt s}, M_{\tt s},\chi_2,\frac{L^{\tt NL}}{ r^{\alpha^{\tt NL}}}\right)f_{R_{\tt s}^{\tt NL}}(r) dz dr ,
\end{align}
where
\begin{align}\label{Xi}
&\Xi\left(x_1,x_2,x_3,x_4\right) \triangleq \notag\\
&\exp\nsp \left(\nsp -2 \pi \lambda_{\tt m}\left[ \int_{x_1}^\infty \nsp \left(1-e^{-\frac{z P_{\tt m}L^{\tt L}}{\rho_1u^{\alpha^{\tt L}}} }\right)\bbP_{\tt m}^{\tt L}(u) u du  +\nsp
\int_{x_2}^\infty \nsp \left(1-e^{-\frac{z P_{\tt m}L^{\tt NL}}{\rho_1 u^{\alpha^{\tt NL}}} }\right)\left(1-\bbP_{\tt m}^{\tt L}(u)\right) u du \right]\right)\notag\\
& \times \exp\nsp \left(\nsp -2 \pi \lambda_{\tt s} \left[\int_{x_3}^\infty \nsp \left(1-e^{-\frac{z P_{\tt s}L^{\tt L}}{\rho_1 u^{\alpha^{\tt L}}} }\right)\bbP_{\tt s}^{\tt L}(u) u du
+\nsp \int_{x_4}^\infty\nsp \left(1-e^{-\frac{z P_{\tt s}L^{\tt NL} }{\rho_1 u^{\alpha^{\tt NL}}} }\right)\left(1-\bbP_{\tt s}^{\tt L}(u)\right) u du \right]\right),
\end{align}
and
\begin{align}
&\Psi\left(x_1,x_2,x_3,x_4,x_5\right)  \triangleq \notag\\ &\exp\nsp \left(-\frac{zx_1}{\rho_1}\left[x_5-\frac{x_5^2}{x_2\left(x_5\nsp +\nsp x_4\nsp +\nsp \sigma^2/\tau p_p\right)}\right]\right)-\exp\nsp \left(-\frac{z x_1}{\rho_1}\left[x_5+\frac{x_3 x_5^2}{x_2\left(x_5 \nsp +\nsp x_4\nsp +\nsp \sigma^2/\tau p_p\right)}\right]\right),
\end{align}
while $\chi_1 \triangleq 2 \pi \int_{C_{\tt v}}^\infty u \lambda_{\mathcal{N}'_0}(u) \left[L^{\tt L} u^{-\alpha^{\tt L}} \bbP_{\tt m}^{\tt L} (u) + L^{\tt NL} u^{-\alpha^{\tt NL}}\left(1-\bbP_{\tt m}^{\tt L} (u) \right)\right] du$, $\chi_2 \triangleq 2 \pi \int_{C_{\tt v}}^\infty u \lambda_{\mathcal{N}'_0}(u) \left[L^{\tt L} \right. \\ \left.u^{-\alpha^{\tt L}} \bbP_{\tt s}^{\tt L} (u) + L^{\tt NL} u^{-\alpha^{\tt NL}}\left(1-\bbP_{\tt s}^{\tt L} (u) \right)\right] du$, $\mathbb{N}_{\tt m} \triangleq {\mathcal{A}_{\tt m}\left(\lambda_{\tt m} N -1\right)}/{\lambda_{\tt m}} +1$, $\mathbb{N}_{\tt s} \triangleq  {\mathcal{A}_{\tt s}\left(\lambda_{\tt m} N-1\right)}/{\lambda_{\tt s}} +1$, $\rho_1 \triangleq \tilde \mu_1 + \sigma^2$, where
\begin{equation}\label{mu1 approximation final}
\tilde \mu_1 \triangleq \frac{\mathcal{A}_{\tt m}^{\tt L} P_{\tt m} M_{\tt m}\xi_1 }{\mathbb{N}_{\tt m} \left(\nu_1 \nmsp +\nmsp \chi_1 \nmsp +\nmsp \frac{\sigma^2}{\tau p_p} \right) } +  \frac{\mathcal{A}_{\tt m}^{\tt NL} P_{\tt m} M_{\tt m}\xi_1 }{\mathbb{N}_{\tt m} \left(\nu_2 \nmsp +\nmsp  \chi_1 \nmsp +\nmsp \frac{\sigma^2}{\tau p_p} \right) }
 +  \frac{\mathcal{A}_{\tt s}^{\tt L} P_{\tt s} M_{\tt s}\xi_2  }{\mathbb{N}_{\tt s} \left(\nu_3 \nmsp +\nmsp  \chi_2 \nmsp +\nmsp \frac{\sigma^2}{\tau p_p} \right) }+ \frac{\mathcal{A}_{\tt s}^{\tt NL} P_{\tt s} M_{\tt s}\xi_2  }{\mathbb{N}_{\tt s} \left(\nu_4\nmsp +\nmsp  \chi_2 \nmsp +\nmsp \frac{\sigma^2}{\tau p_p} \right) },
\end{equation}
with $\xi_1 \triangleq 2\pi \lambda_{\tt m} \int_{C_{\tt v}}^\infty u\left[\left(L^{\tt L}\right)^2 u^{-2\alpha^{\tt L}} \bbP_{\tt m}^{\tt L}(u) + \left(L^{\tt NL}\right)^2 u^{-2\alpha^{\tt NL}}\left(1- \bbP_{\tt m}^{\tt L}(u)\right)\right]du$, and $\xi_2 \triangleq 2\pi \lambda_{\tt m} \int_{C_{\tt v}}^\infty u\\ \left[\left(L^{\tt L}\right)^2 u^{-2\alpha^{\tt L}} \bbP_{\tt s}^{\tt L}(u) + \left(L^{\tt NL}\right)^2 u^{-2\alpha^{\tt NL}}\left(1- \bbP_{\tt s}^{\tt L}(u)\right)\right]du$, as well as $\nu_1 \triangleq \int_0^\infty L^{\tt L} u^{-\alpha^{\tt L}} f_{R_{\tt m}^{\tt L}} (u) du$, $\nu_2 \triangleq \int_0^\infty L^{\tt NL} u^{-\alpha^{\tt NL}} f_{R_{\tt m}^{\tt NL}} (u) di$, $\nu_3 \triangleq \int_0^\infty L^{\tt L} u^{-\alpha^{\tt L}} f_{R_{\tt s}^{\tt L}} (u) du$, and $\nu_4 \triangleq \int_0^\infty L^{\tt NL} u^{-\alpha^{\tt NL}} f_{R_{\tt s}^{\tt NL}} (u) du$.
\end{theorem}
\begin{IEEEproof}
See Appendix \ref{proof of theorem 3}.
\end{IEEEproof}

From Theorem \ref{theorem 3}, we can observe the effects of $\lambda_{\tt s}$ and $M_{\tt m}$ on the achievable downlink rate as described in following remarks. Note that increasing $\lambda_{\tt s}$ means deploying more small cells, and increasing $M_{\tt m}$ means expanding antenna array at the MBS.

\begin{remark}\label{remark 2}
It's not easy to determine the behavior of \eqref{rate approximation} with respect to $M_{\tt m}$ directly, but we can facilitate it with the aid of \eqref{rate mL 2 temp1} as well as numerical results. From \eqref{rate mL 2 temp1}, we know that the downlink rate in \eqref{rate approximation} is a monotonic function of $M_{\tt m}$. Meanwhile, Fig. \ref{fig changeWithMm} reveals that the downlink rate grows as the increment of $M_{\tt m}$. Therefore, it can be deduced that the downlink rate monotonically increases with $M_{\tt m}$ at all the feasible region, which also coincides with our general intuition. When $M_{\tt m} \to \infty$, the saturation of the downlink rate to a constant value can be directly observed form \eqref{rate approximation}.
\end{remark}

\begin{remark}\label{remark 3}
When $\lambda_{\tt s}$ increases, more users will be associated with SCBs, so $\mathcal{A}_{\tt s}^{\tt L}$ and $\mathcal{A}_{\tt s}^{\tt NL}$ grow, and $\mathcal{A}_{\tt m}^{\tt L}$ and $\mathcal{A}_{\tt m}^{\tt NL}$ reduce, which further result in the impact of $M_{\tt m}$ on downlink rate decays. Larger $\lambda_{\tt s}$ can shorten the transmission distance and improve the achievable rate, but when $\lambda_{\tt s}$ exceeds a critical threshold, the aggravation of interferences due to switches from NLoS to LoS will instead impair the rate performance.
\end{remark}

Equation \eqref{rate approximation} quantifies how all the key features of an HCN, i.e., LoS/NLoS transmissions, interference, and deployment strategy affect the achievable downlink rate. The validation of analysis as well as several numerical results based on \eqref{rate approximation} will be shown in Section~\ref{sec: numerical results} to give more practical insights into the design of HCN.

\section{Numerical Results}\label{sec: numerical results}
In this section, we validate the accuracy of our analysis through simulations and evaluate the performance of the HCN via numerical results. In particular, we compare the performance of downlink rate achieved by massive MIMO and small cells. Then, the optimal network configuration parameters are provided as useful guidance for practical implementations. Finally, we explore the antenna allocation between centralized MBSs and distributed SCBs.

From a practical perspective, we use the linear LoS probability function adopted in 3GPP \cite{ahg03} for both the MBS and SCB path losses, which is given by
\begin{equation}\label{3GPP LoS Prob}
\bbP_{\tt m}^{\tt L}(r) = \bbP_{\tt s}^{\tt L}(r)= \left\{
\begin{array}{ll}
1-r/d_{\tt L}, & 0<r \le d_{\tt L}. \\
0,& r > d_{\tt L}.\\
\end{array}\right.
\end{equation}
According to \cite{ahg03,3GPP12}, parameters used in our simulation are set as follows: $d_{\tt L}=0.3$ km, $L^{\tt L}=10^{-10.38}$, $L^{\tt NL}=10^{-14.54}$, $\alpha^{\tt L}=2.09$, $\alpha^{\tt NL}=3.75$, $P_{\tt m}=53$ dBm, $P_{\tt s}=33$ dBm, and $\sigma^2=-104$ dBm. The uplink pilot power is $p_p = 24$ dBm, and the MBS density is $\lambda_{\tt m} = 1$ BSs/km$^2$.

\subsection{Validation of Analytical Results}
\begin{figure}[t!]
\centering
\includegraphics[width=88mm,height=78mm]{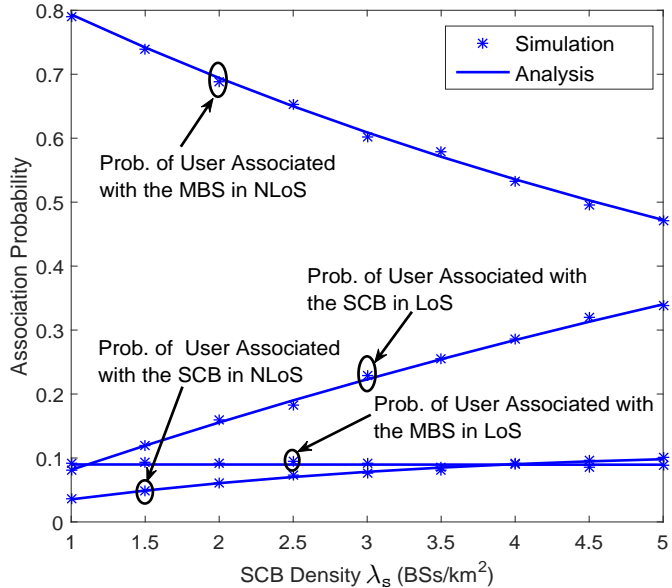}
\caption{User association probability vs. SCB density $\lambda_{\tt s}$, where $B=1$.}\label{fig compare_assoProb}
\end{figure}

 In Fig. \ref{fig compare_assoProb}, the simulated user association probability is compared with our analytical results in Theorem \ref{theorem 1}. Clearly, we can see perfect agreement between the simulated and analytical values, which justifies the accuracy of our calculations. When $\lambda_{\tt s}$ increases, as expected, the probability of user association with the MBS reduces and that with the SCB grows. Moreover, all the probabilities change with $\lambda_{\tt s}$ almost linearly, which coincides with the conclusion in Remark \ref{remark 1} that the average number of users associated with each SCB remains unchanged as $\lambda_{\tt s}$ increases. Fig. \ref{fig compare_assoProb} also reveals that the macro cell users are more likely in NLoS transmissions while small cell uses are more likely in LoS transmissions. This is because the sparse deployment of MBSs leads to a large serving distance between users and their associated MBSs, which results in rare LoS paths, and in contrary, the dense SCBs make the LoS connection to a user more often.

\begin{figure}[!t]
\centering
\includegraphics[width=88mm,height=78mm]{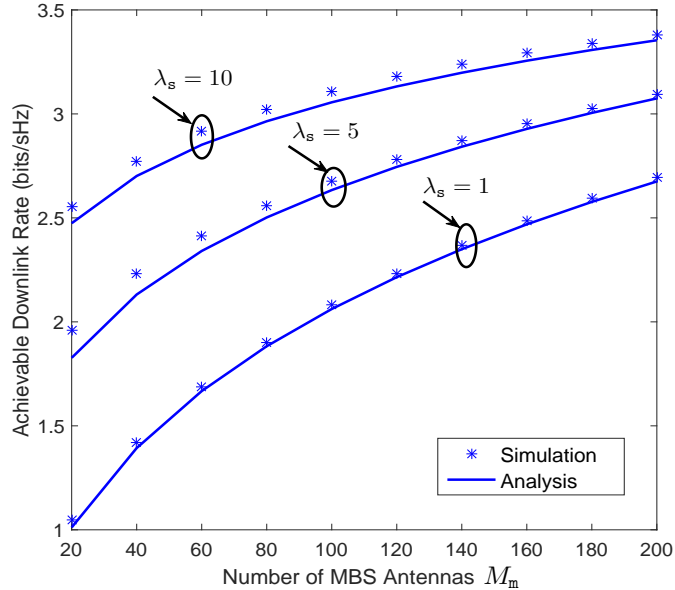}
\caption{Achievable downlink rate vs. number of MBS antennas, where $N=10$, $M_{\tt s}=5$ and $B=1$.}\label{fig compare_withM}
\end{figure}

In Fig. \ref{fig compare_withM}, the simulated achievable downlink rate is compared with the analytical approximation in \eqref{rate approximation} under different values of $M_{\tt m}$ and $\lambda_{\tt s}$. We can see that the analytical approximation and simulation results fairly well match and follow the same trend, thus verifies the accuracy of Theorem \ref{theorem 3}.

Due to the tightness between the simulations and analysis, we will use the latter for our following investigations. Note that the number of BS antennas should be larger than the number of users it serves. Hence, $M_{\tt m}$ is set to be larger than $N$, and $M_{\tt s}$ is set according to the average number of users associated with each SCB as given in Corollary \ref{corollary 1}. From Fig. \ref{fig compare_assoProb}, we can get the curve slop by line fitting and obtain that $\mathcal{A}_{\tt s} \approx 0.08 \lambda_{\tt s}$. Therefore, the average number of users associated with each SCB is $\mathcal{N}_{\tt s} \approx 0.08 \lambda_{\tt m} N$.

\subsection{Comparison between Massive MIMO and Small Cells}
\begin{figure}[!t]
\centering
\includegraphics[width=88mm,height=78mm]{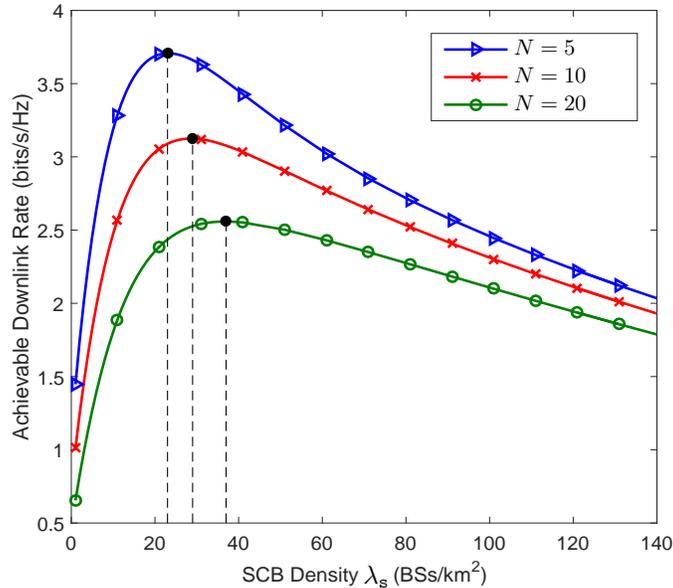}
\caption{Achievable downlink rate vs. SCB density $\lambda_{\tt s}$, where $M_{\tt s}=5$, $M_{\tt m}=20$ and $B=1$.}\label{fig changeWithLambdaSCB}
\end{figure}

\begin{figure}[!t]
\centering
\includegraphics[width=88mm,height=78mm]{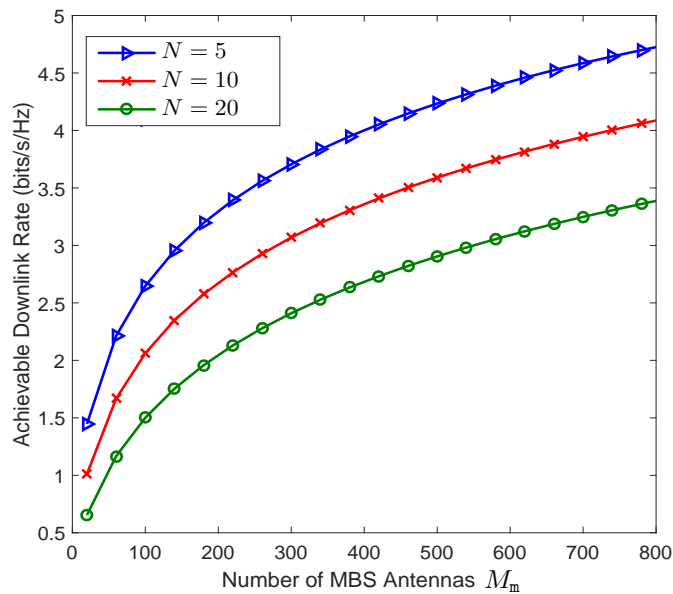}
\caption{Achievable downlink rate vs. number of MBS antennas $M_{\tt m}$, where $\lambda_{\tt s}=1$ BSs/km$^2$, $M_{\tt s}=5$ and $B=1$.}\label{fig changeWithMm}
\end{figure}

In this subsection, we aim to compare the performance of massive MIMO and small cells. In particular,
 Fig. \ref{fig changeWithLambdaSCB} reveals the effect of increasing $\lambda_{\tt s}$ with fixed $M_{\tt m}$, while Fig. \ref{fig changeWithMm} shows the effect of increasing $M_{\tt m}$ with fixed $\lambda_{\tt s}$. For fairness, we set the same starting configuration for these two cases, where the MBS antenna array is $M_{\tt m}=20$ and the SCB density is $\lambda_{\tt s}=1$ BSs/km$^2$.

\begin{figure}[!t]
\centering
\includegraphics[width=85mm,height=78mm]{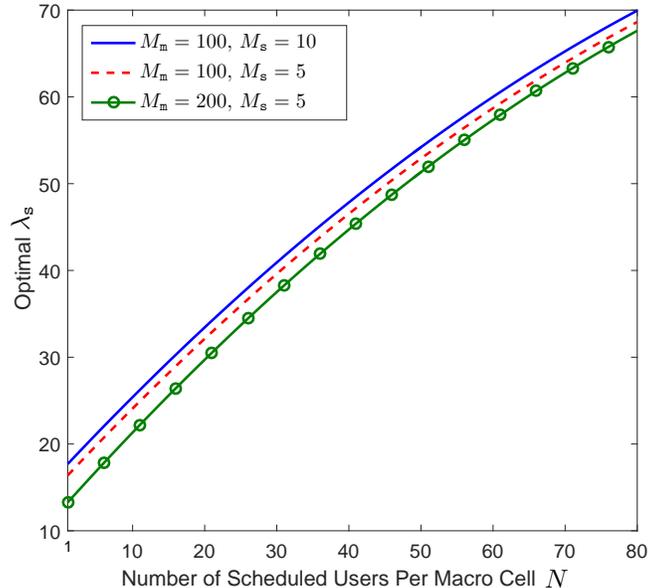}
\caption{The optimal $\lambda_{\tt s}$ vs. number of scheduled users per macro cell $N$, where $B=1$.}\label{fig optimalLambdaSCB}
\end{figure}

 Fig. \ref{fig changeWithLambdaSCB} shows that the downlink rate burgeons with the increment of $\lambda_{\tt s}$ until reaching a critical threshold, after which the expectancy for further improvement breaks into a slow decline.
This is because when $\lambda_{\tt s}$ is small, the network can benefit a lot from the small cell densification due to the reduced distance between transceivers. However, when $\lambda_{\tt s}$ becomes large, more and more interference paths switch from NLoS to LoS, resulting in an aggregated interference that significantly impair the rate performance. This observation is consistent with the conclusion in \cite{ding16} where the set up is with single antenna BSs. The critical SCB density threshold, i.e., the optimal $\lambda_{\tt s}$ that maximizes the downlink rate, is marked out by black dots. Note that as the scheduled user number $N$ grows, the optimal $\lambda_{\tt s}$ also grows, since more small cells are needed in a more crowded environment.
The optimal $\lambda_{\tt s}$ for different $N$ is summarized in Fig. \ref{fig optimalLambdaSCB}.

Fig. \ref{fig changeWithMm} shows that the donwlink rate increases monotonically with respect to $M_{\tt m}$, but it cannot grow without bound and saturates to a constant value limited by pilot contamination. It can be also observed that the smaller $N$ leads to a better rate performance, because we are considering the achievable rate per user, and less number of users gives less interferences.

By comparing Figs. \ref{fig changeWithLambdaSCB} and \ref{fig changeWithMm}, we can see that adding small cells into a sparse network is more effective in boosting up the achievable rate than expanding antenna arrays at MBSs. Densification is remarkably beneficial when $\lambda_{\tt s}$ is low, where the rate can be enhanced almost linearly. For example, when $N=10$, we observe that increasing $\lambda_{\tt s}$ by $10$ can bring almost $100\%$ rate gain. That means, on average, adding $10$ small cells into per macro cell can double the downlink rate. In contrast, for the massive MIMO system in Fig. \ref{fig changeWithMm}, we need to add more than $100$ antennas to have the same rate gain. However, the peak rate obtained from small cell densification is lower than that from antenna array extension. Besides, as more number of antennas can tolerate more users in a single cell, massive MIMO achieves better sum rate than small cell technique.
 Therefore, we conclude that the deploying small cells can improve the achievable rate fast and effectively, but it will stop benefiting and further impair the network capacity when the SCB density exceeds a critical threshold. On the contrary, though the massive MIMO technique improves the system performance slower, but expanding antenna size can always benefit the network capacity, and the maximum rate with large $M_{\tt m}$ is greater than that obtained from small cell deployment. In summary, if the rate demand is low, deploying smalls cell is preferred due to its rapid rate gain; but if the rate requirement is high, the massive MIMO technique is more preferable due to the higher achievable rate it provides.

\subsection{Optimal Network Configuration}
In this subsection, we are interested in obtaining the optimal SCB density $\lambda_{\tt s}$ that can maximize the achievable downlink rate as shown in Fig. \ref{fig changeWithLambdaSCB}. Moreover, with abandoned antennas available at MBSs, the biasing policy should be adjusted to fully exploit the excessive degrees-of-freedom. Therefore, we also investigate the impact of biasing factor in this subsection.

Fig. \ref{fig changeWithLambdaSCB} illustrates how the optimal $\lambda_{\tt s}$ varies with the number of users scheduled by per MBS, $N$, under different BS antenna numbers.
 We can see that the optimal $\lambda_{\tt s}$ increases monotonically with $N$, since the most effective way to offload traffic caused by crowed users is deploying more SCBs. We also find that the optimal $\lambda_{\tt s}$ increases $M_{\tt s}$ and decreases with $M_{\tt m}$, which coincides with general intuition as a powerful MBS with large antenna array can provide sufficient data rate for users in its coverage thus requires less SCB deployment, while SCBs with more antennas are desirable for rate enhancement thus the optimal $\lambda_{\tt s}$ increases.

\begin{figure}[!t]
\centering
\includegraphics[width=88mm,height=78mm]{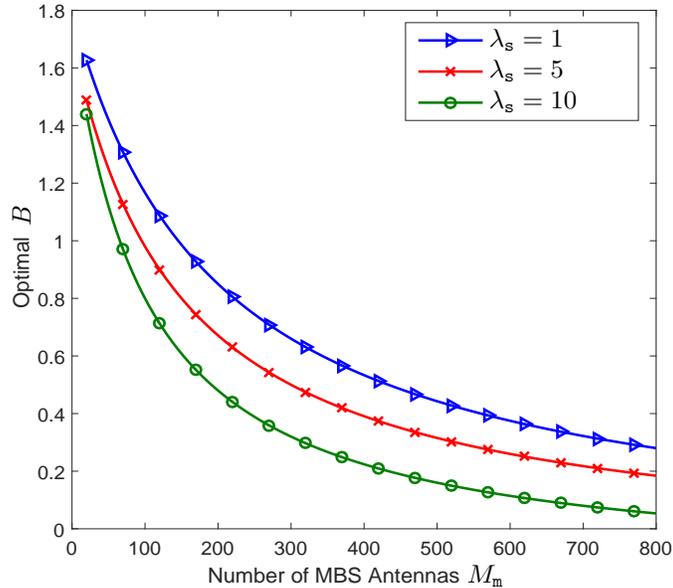}
\caption{The optimal $B$ vs. number of MBS antennas $M_{\tt m}$, where $M_{\tt s}=5$ and $N=10$.}\label{fig optimalB}
\end{figure}

 Fig. \ref{fig optimalB} presents the optimal SCB bias factor $B$ under different MBS antenna number $M_{\tt m}$. Clearly, the optimal $B$ decreases as $M_{\tt m}$ grows, since larger $M_{\tt m}$ can bring more significant rate gain thus a smaller $B$ is desired to push more users associated with MBSs and benefit from the vast degrees-of-freedom.
We also note that the optimal $B$ also decreases with increment of $\lambda_{\tt s}$ due to a similar argument.

All the above results can serve as useful guidance for practical network design.

\subsection{Comparison of Distributed and Centralized Antennas}
In this subsection, we explore the antenna resource allocation between the MBS and SCBs under fixed amount of antenna budget, i.e., we aim to investigate how many antennas should be allocated to the MBS and SCBs, respectively, such that the system performance can be maximized. Let $M$ be the total number of antennas per macro cell. Then, on average, $M = \lambda_{\tt s} M_{\tt s} + M_{\tt m}$. The proportion of antennas assigned to SCBs per macro cell is defined as
$\varpi \triangleq {\lambda_{\tt s} M_{\tt s}}/{M}$.

\begin{figure}[!t]
\centering
\includegraphics[width=88mm,height=78mm]{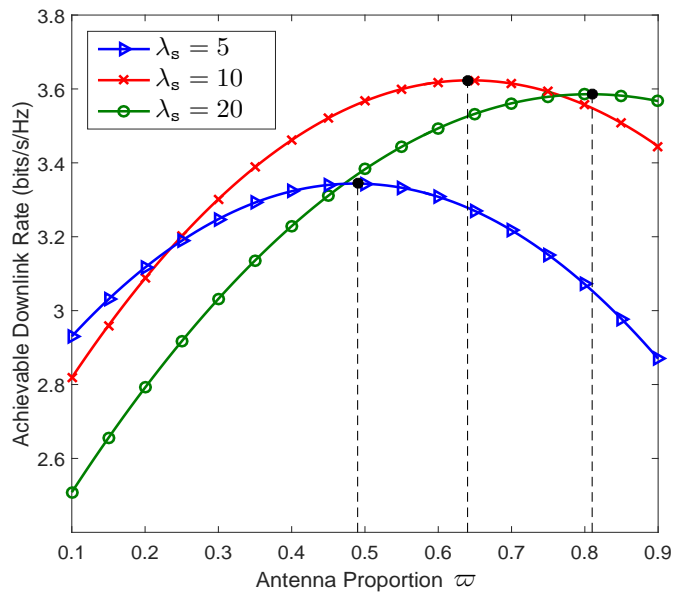}
\caption{Achievable downlink rate vs. the antenna proportion $\varpi$, where $M=200$ and $N=10$.}\label{fig changeWithRatio}
\end{figure}

 Fig. \ref{fig changeWithRatio} shows the achievable downlink rate as a function of $\varpi$ for a fixed $M=200$. We can see that the rate first increases as $\varpi$ grows, which indicates that spare certain amount of centralized MBS antennas to the distributed SCBs can improve the system performance. However, when $\varpi$ exceeds a critical value, the rate begins to abate, which means assigning too many antennas to SCBs will leave MBS without enough spatial diversity and impair the network capacity. We also note that the optimal $\varpi$ that maximizes the downlink rate, as marked out by black dots, increases with $\lambda_{\tt s}$.

\begin{figure}[!t]
\centering
\includegraphics[width=88mm,height=78mm]{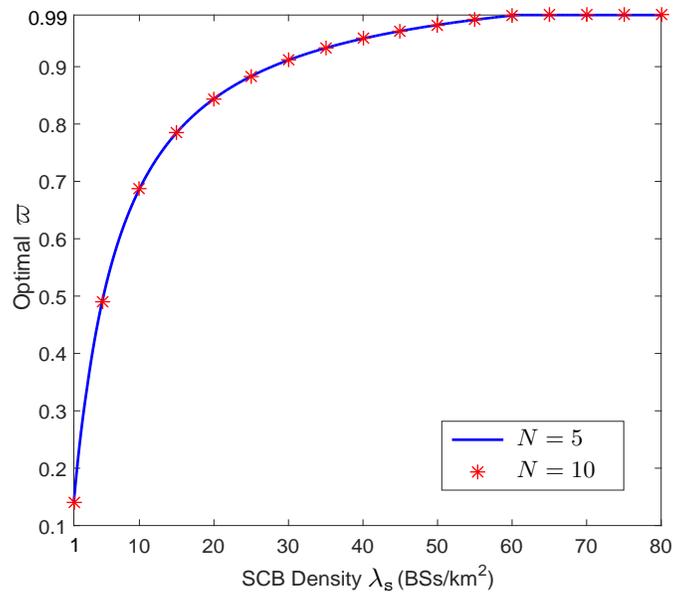}
\caption{The optimal $\varpi$ vs. SCB density $\lambda_{\tt s}$, where $M=200$.}\label{fig optimalRatio}
\end{figure}

More precise information about the optimal $\varpi$ is summarized in Fig. \ref{fig optimalRatio}.
We find that the optimal $\varpi$ grows with SCB density $\lambda_{\tt s}$ while remains unchanged for different $N$. This is because more antennas need to be assigned to SCBs when their density increases, while the user density has no impact on the optimal portion.
Particularly, the optimal $\varpi$ is limited by an upper bounded smaller than $1$, since we cannot allocate all antennas to SCBs as the MBS needs to schedule users and assign pilots before data transmission.
The practical antenna allocation can get useful references from this figure while combining with hardware constraints.

\section{Conclusion}\label{sec: conclusion}
In this paper, we have developed a framework for downlink HCN that consists of randomly distributed MBSs and SCBs with multiple antennas, and the LoS and NLoS transmissions are differentiated. Using stochastic geometry, we have derived a tight approximation of achievable rates to compare the performance between densifying small cells and expanding BS antenna arrays. Interestingly, we have found that adding small cells into the network is more effective in boosting up the achievable rate than expanding antenna arrays at MBS. However, when the small cell density exceeds a critical threshold value, the spatial densification stops benefiting and further impairs the network capacity. In contrast, expanding BS antenna array can always improve the capacity until reaching an upper bound caused by pilot contamination, and this upper bound is larger than the peak rate obtained from deployment of small cells. Therefore, for low rate requirements, the small cell is preferred due to its sheer rate gain; but for a higher rate requirements, the massive MIMO is preferred due to better achievable rate.
The optimal SCB density is also presented as a guidance for practical small cell deployment. Moreover, we have found that allocating part of antennas to the distributed SCBs is better than centralizing all antennas at the macro BS, and the optimal allocation proportion has also been provided for practical configuration reference. In conclusion, this work has provided a further understanding on how to leverage small cells and massive MIMO in future HCNs deployment.

\appendices

\section{Proof of Theorem \ref{theorem 1}}\label{proof of theorem 1}
In general, the typical user can receive four types of transmit powers, i.e., from the MBS through LoS and NLoS path, and from SCB through LoS and NLoS path, respectively. The typical user is associated with the MBS in a LoS path means that this received power is higher than other three cases, which can be formulated from \eqref{brp macro} and \eqref{brp scb} as follows
\begin{align}\label{Prob EmL appendix}
\bbP\left[E_{\tt m}^{\tt L}\right]
&=\bbP\left[ \left\{ \frac{ P_{\tt m} L^{\tt L} }{ \left(S_{\tt m}^{\tt L}\right)^{\alpha^{\tt L}} } > \frac{P_{\tt m} L^{\tt NL}}{  \left(S_{\tt m}^{\tt NL}\right)^{\alpha^{\tt NL}} }  \right\} \bigcap \left\{ \frac{P_{\tt m} L^{\tt L}}{\left(S_{\tt m}^{\tt L}\right)^{\alpha^{\tt L}}}  > \frac{B P_{\tt s} L^{\tt L}}{ \left(S_{\tt s}^{{\tt L}}\right)^{\alpha^{\tt L}} }  \right\} \bigcap \left\{ \frac{P_{\tt m} L^{\tt L}}{\left(S_{\tt m}^{\tt L}\right)^{\alpha^{\tt L}}} > \frac{B P_{\tt s} L^{\tt NL}}{ \left(S_{\tt s}^{{\tt NL}}\right)^{\alpha^{\tt NL}} }  \right\}  \right]\notag\\
&=\! \int_0^\infty\!\! \bbP\!\left[S_{\tt m}^{\tt NL} \!>\! k_1 r^{\frac{\alpha^{\tt L}}{\alpha^{\tt NL}}}|S_{\tt m}^{\tt L} \!=\! r\right] \! \cdot \! \bbP\!\left[S_{\tt s}^{{\tt L}} \!>\! k_2 r|S_{\tt m}^{\tt L} = r\right] \! \cdot \! \bbP\!\left[S_{\tt s}^{{\tt NL}} \!>\! k_1 k_3 r^{\frac{\alpha^{\tt L}}{\alpha^{\tt NL}}}| S_{\tt m}^{\tt L} \!=\! r\right] \! \cdot \! f_{S_{\tt m}^{\tt L}} (r) dr.
\end{align}

With results from void probability, the first term in the integral of \eqref{Prob EmL appendix} can be calculated as follows
\begin{align}\label{t1}
 \bbP\left[S_{\tt m}^{\tt NL} > k_1 r^{\frac{\alpha^{\tt L}}{\alpha^{\tt NL}}}|S_{\tt m}^{\tt L} = r\right]
&=  \bbP\left[\text{ No BS in } \Phi_{\tt m}^{\tt NL} \text{ is closer than } k_1 r^{\frac{\alpha^{\tt L}}{\alpha^{\tt NL}}}\right]\notag\\
&= \exp\left(-\int_0^{k_1 r^{\frac{\alpha^{\tt L}}{\alpha^{\tt NL}}}} \lambda_{\tt m}(1-\bbP_{\tt m}^{\tt L}(u)) 2 \pi u du\right).
\end{align}
Similarly, we can derive the second and third terms in \eqref{Prob EmL appendix} respectively as follows
\begin{align}\label{t2}
\bbP\left[S_{\tt s}^{{\tt L}}> k_2 r|S_{\tt m}^{\tt L} = r\right] &=\exp\left(-\int_0^{k_2 r} \lambda_{\tt s}\bbP_{\tt s}^{\tt L}(u) 2 \pi u du\right),\notag\\
\bbP\left[S_{\tt s}^{{\tt NL}}> k_1 k_3 r^\frac{\alpha^{\tt L}}{\alpha^{\tt NL}}|S_{\tt m}^{\tt L} = r\right] &=\exp\left(-\int_0^{k_1 k_3 r^\frac{\alpha^{\tt L}}{\alpha^{\tt NL}}} \lambda_{\tt s}(1-\bbP_{\tt s}^{\tt L}(u)) 2 \pi u du\right).
\end{align}
Moreover, the pdf of $S_{\tt m}^{\tt L}$ can be obtained as
\begin{equation}\label{f Sml}
f_{S_{\tt m}^{\tt L}}(r)= \frac{d(1-\bbP\left[S_{\tt m}^{\tt L}>r\right])}{dr}=\exp\left(-\int_0^r \lambda_{\tt m} \bbP_{\tt m}^{\tt L} (u) 2 \pi u du \right)  \bbP_{\tt m}^{\tt L} (r) 2 \pi r \lambda_{\tt m}.
\end{equation}
Therefore, \eqref{Prob EmL} follows from the substitution of \eqref{t1}---\eqref{f Sml} into \eqref{Prob EmL appendix}.
By performing the same procedure, we can obtain the probabilities $\bbP\left[E_{\tt m}^{\tt NL}\right]$, $\bbP\left[E_{\tt s}^{\tt L}\right]$, and $\bbP\left[E_{\tt s}^{\tt NL}\right]$.

\section{Proof of Corollary \ref{corollary 1}}\label{proof of corollary 1}
Since each MBS randomly schedules $N$ users into its cell, the density of scheduled users are $\lambda_{\tt m} N$. Let $\mathcal{S}$ be the area of the entire network. Then, the total number of scheduled users in the network is $\lambda_{\tt m} N \mathcal{S} $, where, on average, $\mathcal{A}_{\tt m} \lambda_{\tt m} N \mathcal{S} $ users are associated with the MBS, and $\mathcal{A}_{\tt s} \lambda_{\tt m} N \mathcal{S}$ users are associated with the SCB. Therefore, considering that the total number of MBSs in the network is $\lambda_{\tt m} \mathcal{S}$ and the total number of SCBs is $\lambda_{\tt s} \mathcal{S}$, we can get average number of users associated with each MBS as
\begin{equation}\label{average number of user MBS}
\mathcal{N}_{\tt m} = \frac{\mathcal{A}_{\tt m}\lambda_{\tt m} N \mathcal{S}}{\lambda_{\tt m} \mathcal{S} }= \mathcal{A}_{\tt m} N,
\end{equation}
and the average number of users associated with each SCB as
\begin{equation}\label{average number of user SCB}
\mathcal{N}_{\tt s} = \frac{\mathcal{A}_{\tt s}\lambda_{\tt m} N \mathcal{S}}{\lambda_{\tt s} \mathcal{S} }= \frac{\mathcal{A}_{\tt s}\lambda_{\tt m} N }{\lambda_{\tt s} }.
\end{equation}

\section{Proof of Theorem \ref{theorem 2}}\label{proof of theorem 2}
In order to derive the pdf of $R_{\tt m}^{\tt L}$, we first investigate its complementary cumulative distribution function as $\bar F_{R_{\tt m}^{\tt L}}(r) = \bbP\left[R_{\tt m}^{\tt L} > r\right]$. The event $R_{\tt m}^{\tt L} > r$ is equivalent to that $S_{\tt m}^{\tt L} > r$ given that the typical user is associated with the MBS in a LoS path, i.e.,
\begin{equation}
\bbP\left[R_{\tt m}^{\tt L} > r\right]=\bbP\left[S_{\tt m}^{\tt L}> r | E_{\tt m}^{\tt L}\right] = \frac{\bbP\left[S_{\tt m}^{\tt L}> r , E_{\tt m}^{\tt L}\right]}{ \bbP\left[E_{\tt m}^{\tt L}\right]}.
\end{equation}
A similar argument as in \eqref{Prob EmL appendix} leads us to the following calulation
\begin{align}\label{t21}
\bbP\left[S_{\tt m}^{\tt L}>r, E_{\tt m}^{\tt L}\right] & =\int_r^\infty  \bbP\left[P_{\tt m} L^{\tt L} \left(S_{\tt m}^{\tt L}\right)^{-\alpha^{\tt L}}>P_{\tt m} L^{\tt NL} \left(S_{\tt m}^{\tt NL}\right)^{-\alpha^{\tt NL}}| S_{\tt m}^{\tt L} = x\right]\notag\\
&~~~~~~\times \bbP\left[P_{\tt m} L^{\tt L} \left(S_{\tt m}^{\tt L}\right)^{-\alpha^{\tt L}}>P_{\tt s} L^{\tt L} \left(S_{\tt s}^{{\tt L}}\right)^{-\alpha^{\tt L}}B | S_{\tt m}^{\tt L} = x\right]\notag\\
&~~~~~~\times \bbP\left[P_{\tt m} L^{\tt L} \left(S_{\tt m}^{\tt L}\right)^{-\alpha^{\tt L}}>P_{\tt s} L^{\tt NL} \left(S_{\tt s}^{{\tt NL}}\right)^{-\alpha^{\tt NL}}B| S_{\tt m}^{\tt L} = x\right]f_{S_{\tt m}^{\tt L}}(x) dx,
\end{align}
By using results in \eqref{t1}--\eqref{f Sml}, we have
\begin{align}\label{t22}
\bbP\left[R_{\tt m}^{\tt L}>r\right] &= \frac{2 \pi \lambda_{\tt m}}{\mathcal{A}_{\tt m}^{\tt L}}  \int_r^\infty x \bbP_{\tt m}^{\tt L}(x) \zeta_1\left(x,k_1 x^\frac{\alpha^{\tt L}}{\alpha^{\tt NL}}\right)\zeta_2\left(k_2 x, k_1 k_3 x^\frac{\alpha^{\tt L}}{\alpha^{\tt NL}}\right) dx.
\end{align}
The pdf of $R_{\tt m}^{\tt L}$ is then follows from taking derivative of $1-\bar F_{R_{\tt m}^{\tt L}}(r)$ with respect to $r$.
The pdf of $R_{\tt m}^{\tt NL}$, $R_{\tt s}^{\tt L}$ and $R_{\tt s}^{\tt NL}$ can be obtained from the same procedure, which are omitted due to space limits.

\section{Proof of Theorem \ref{theorem 3}}\label{proof of theorem 3}

We take the derivation of ${\tt SINR}_{\tt m}^{\tt L}$ for example. With the maximal-ratio-transmission precoder,
when the typical user is associated with the MBS $0$ in LoS path, from \eqref{sinr macro}, we can get
\begin{align}\label{rate mL jensen}
\bbE\left\{\log_2\left(1+{\tt SINR}_{\tt m}^{\tt L}\right)\right\}
 &  =   \bbE\left\{\log_2 \left(1+\frac{P_{\tt m} \kappa_0^{{(\tt m)}2} \left|\bhg_{000}^{{(\tt m)}H}\bhg^{(\tt m)}_{000}\right|^2}{\left({\tt I}_{\tt m}^{\tt MRT}+ \sigma^2 \right)\bbE_{\bh}\left\{\left\|\bhg_{000}^{(\tt m)}\right\|^2\right\}}\right)\right\}\notag\\
& \mathop \approx \limits^{(a)} \bbE_{\bs \varphi}\left\{ \log_2 \left(1+\frac{P_{\tt m}\kappa_0^{{(\tt m)}2} \bbE_\bh\left\{\left|\bhg_{000}^{{(\tt m)}H}\bhg^{(\tt m)}_{000}\right|^2\right\}}{\left(\bbE_{\bh}\left\{{\tt I}_{\tt m}^{\tt MRT}\right\}+ \sigma^2 \right)\bbE_{\bh}\left\{\left\|\bhg_{000}^{(\tt m)}\right\|^2\right\}}\right)\right\},
\end{align}
where $\bbE_{\bs\varphi}$ denotes the average over path loss, $(b)$ follows from the approximation in \cite[Lemma 1]{qi14}, and ${\tt I}_{\tt m}^{\tt MRT}$ is given as
\begin{multline}\label{I mL}
{\tt I}_{\tt m}^{\tt MRT} =
P_{\tt m}\kappa_0^{{(\tt m)}2} \nsp \nsp \nsp \nsp   \sum\limits_{U_{n0}\in \mathcal{U}^{\tt m}_{0} \backslash U_{00}} \frac{\left|\bhg_{000}^{{(\tt m)}H} \bhg^{(\tt m)}_{0n0} \right|^2}{\bbE_{\bh}\left\{\left\|\bhg_{0n0}^{(\tt m)}\right\|^2\right\}} +P_{\tt m} \kappa_0^{{(\tt m)}2} \nsp \sum\limits_{U_{n0}\in \mathcal{U}^{\tt m}_{0}}\frac{\left|\btg_{000}^{{(\tt m)}H} \bhg^{(\tt m)}_{0n0} \right|^2}{\bbE_{\bh}\left\{\left\|\bhg_{0n0}^{(\tt m)}\right\|^2\right\}}
\\ + P_{\tt m} \nsp \sum\limits_{l \in \Phi_{\tt m}\backslash 0}\sum\limits_{U_{nl}\in \mathcal{U}^{\tt m}_{l}}\nsp \nsp \kappa_l^{{(\tt m)}2} \frac{\left|\bg_{l00}^{{(\tt m)}H}\bhg^{(\tt m)}_{lnl}\right|^2}{\bbE_{\bh}\left\{\left\|\bhg_{lnl}^{(\tt m)}\right\|^2\right\}} \nsp +  P_{\tt s} \sum\limits_{j \in \Phi_{\tt s}}\sum\limits_{l \in \Phi_{\tt m}} \sum\limits_{U_{nl}\in \mathcal{U}^{\tt s}_l}\kappa_j^{{(\tt s)}2} \frac{\left|\bg_{j00}^{{(\tt s)}H}\bhg^{(\tt s)}_{jnl}\right|^2}{\bbE_{\bh}\left\{\left\|\bhg_{jnl}^{(\tt s)}\right\|^2\right\}}.
\end{multline}
Next, we calculate the expectations in \eqref{rate mL jensen} separately.

From \eqref{gini estimate} and \eqref{gjni estimate}, we know the variance of elements in $\bhg_{lnl}^{(\tt m)}$ and $\bhg_{jnl}^{(\tt s)}$ are $\varphi_{lnl}^{(\tt m)} \eta_{lnl}^{(\tt m)}$ and $\varphi_{jnl}^{(\tt s)} \eta_{jnl}^{(\tt s)}$, respectively. Hence, we have
\begin{equation}\label{mean g000 g000}
\bbE_{\bh}\left\{\left\|\bhg_{lnl}^{(\tt m)}\right\|^2\right\}=M_{\tt m} \varphi_{lnl}^{(\tt m)} \eta_{lnl}^{(\tt m)},~~~
\bbE_{\bh}\left\{\left\|\bhg_{jnl}^{(\tt s)}\right\|^2\right\}=M_{\tt m} \varphi_{jnl}^{(\tt s)} \eta_{jnl}^{(\tt s)}.
\end{equation}
Moreover,
\begin{align}\label{mean square g000 g000}
\bbE_\bh\left\{\left|\bhg_{000}^{{(\tt m)}H}\bhg^{(\tt m)}_{000}\right|^2\right\}
= &\bbE_\bh\left\{\left(\sum\nolimits_{m=1}^{M_{\tt m}}\hat g_{m000}^{{(\tt m)}*}\hat g_{m000}^{(\tt m)}\right)\left(\sum\nolimits_{m=1}^{M_{\tt m}}\hat g_{m000}^{(\tt m)}\hat g_{m000}^{{(\tt m)}*}\right)\right\}\notag\\
= &\bbE_\bh\left\{\sum_{m_1=1}^{M_{\tt m}}\sum_{m_2\ne m_1}^{M_{\tt m}}\left|\hat g_{m_1000}^{{(\tt m)}*}\right|^2 \left|\hat g_{m_2000}^{{(\tt m)}}\right|^2\right\} + \bbE_\bh\left\{\sum_{m=1}^{M_{\tt m}}\left| \hat g_{m000}^{{(\tt m)}*}\right|^4 \right\}\notag\\
\mathop = \limits^{(a)} & M_{\tt m}\left(M_{\tt m}-1\right) \eta_{000}^{{(\tt m)}4}\left(\sum\limits_{l \in \Phi_{\tt m}} \varphi_{00l}^{(\tt m)} +\frac{\sigma^2}{\tau p_p}\right)^2\nsp + 2 M_{\tt m}\eta_{000}^{{(\tt m)}4}\left(\sum\limits_{l \in \Phi_{\tt m}} \varphi_{00l}^{(\tt m)} + \frac{\sigma^2}{\tau p_p}\right)^2\notag\\
= & \left(M_{\tt m}^2+M_{\tt m}\right)\varphi_{000}^{{(\tt m)}2}\eta_{000}^{{(\tt m)}2},
\end{align}
where $\hat g_{mlnl}^{(\tt m)}$ is the $m$th element of the channel estimation vector $\bhg_{lnl}^{(\tt m)}$, and $(a)$ is obtained from the substitution of \eqref{gini estimate} as well as some basic algebraic operations.

For $U_{n0}\in \mathcal{U}^{\tt m}_{0} \backslash U_{00}$, we obtain
\begin{align}\label{mean square g000 g0n0}
\bbE_\bh\left\{\left|\bhg_{000}^{{(\tt m)}H}\bhg^{(\tt m)}_{0n0}\right|^2\right\}
\mathop = \limits^{(a)}  \varphi_{000}^{(\tt m)}\eta_{000}^{(\tt m)} \bbE_\bh\left\{\left\|\bhg_{0n0}^{(\tt m)}\right\|^2\right\}
=M_{\tt m} \varphi_{000}^{(\tt m)}\eta_{000}^{(\tt m)} \varphi_{0n0}^{(\tt m)}\eta_{0n0}^{(\tt m)},
\end{align}
where $(a)$ is obtained from the independence of $\bhg_{000}^{(\tt m)}$ and $\bhg_{0n0}^{(\tt m)}$, and for $U_{n0}\in \mathcal{U}^{\tt m}_{0}$, we obtain
\begin{align}\label{mean square tildeg000 g0n0}
\bbE_\bh\left\{\left|\btg_{000}^{{(\tt m)}H}\bhg^{(\tt m)}_{0n0}\right|^2\right\}= M_{\tt m} \varphi_{000}^{(\tt m)}\left(1-\eta_{000}^{(\tt m)}\right) \varphi_{0n0}^{(\tt m)}\eta_{0n0}^{(\tt m)},
\end{align}
which is got from the independence of $\btg_{000}^{(\tt m)}$ and $\bhg_{0n0}^{(\tt m)}$.
When $l \in \Phi_{\tt m}\backslash 0 $, the result is dependent on the association of $U_{0l}$, which leads to the following discussion:
\begin{itemize}
  \item If $U_{0l} \in \mathcal{U}_l^{\tt m}$, i.e., $U_{0l}$ is associated with the MBS, we have
  \begin{align}\label{mean square gl00 gl0l}
  \bbE_\bh\left\{\left|\bg_{l00}^{{(\tt m)}H}\bhg^{(\tt m)}_{l0l}\right|^2\right\}= M_{\tt m}^2 \varphi_{l00}^{{(\tt m)}2}\eta_{l0l}^{{(\tt m)}2}+M_{\tt m}\varphi_{l00}^{(\tt m)}\varphi_{l0l}^{(\tt m)}\eta_{l0l}^{(\tt m)},
  \end{align}
 which is obtained according to the similar procedure as \eqref{mean square g000 g000}. For $U_{nl} \in \mathcal{U}_l^{\tt m}\backslash U_{0l}$,
  \begin{align}\label{mean square gl00 glnl}
  \bbE_\bh\left\{\left|\bg_{l00}^{{(\tt m)}H}\bhg^{(\tt m)}_{lnl}\right|^2\right\}= M_{\tt m}\varphi_{l00}^{(\tt m)}\varphi_{lnl}^{(\tt m)}\eta_{lnl}^{(\tt m)},
  \end{align}
 which is obtained according to the similar procedure as \eqref{mean square g000 g0n0}, and for $U_{nl} \in \mathcal{U}_j^{\tt s}$ $(j \in \Phi_{\tt s})$,
  \begin{align}\label{mean square gj00 gjnl}
  \bbE_\bh\left\{\left|\bg_{j00}^{{(\tt s)}H}\bhg^{(\tt s)}_{jnl}\right|^2\right\}= M_{\tt s}\varphi_{j00}^{(\tt s)}\varphi_{jnl}^{(\tt s)}\eta_{jnl}^{(\tt s)},
  \end{align}
  \item If $U_{0l} \in \mathcal{U}_{q_l}^{\tt s}$, i.e., $U_{0l}$ is associated with the SCB, denoted as $q_l$, we have
  \begin{align}\label{mean square gq00 gq0l}
  \bbE_\bh\left\{\left|\bg_{q_l00}^{{(\tt s)}H}\bhg^{(\tt s)}_{q_l0l}\right|^2\right\}= M_{\tt s}^2 \varphi_{q_l00}^{{(\tt s)}2}\eta_{q_l0l}^{{(\tt s)}2}+M_{\tt s}\varphi_{q_l00}^{(\tt s)}\varphi_{q_l0l}^{(\tt s)}\eta_{q_l0l}^{(\tt s)}.
  \end{align}
  For $U_{nl} \in \mathcal{U}_{q_l}^{\tt s} \backslash U_{0l}$,
  \begin{align}\label{mean square gq00 gqnl}
  \bbE_\bh\left\{\left|\bg_{q_l00}^{{(\tt s)}H}\bhg^{(\tt s)}_{q_lnl}\right|^2\right\}= M_{\tt s}\varphi_{q_l00}^{(\tt s)}\varphi_{q_lnl}^{(\tt s)}\eta_{q_lnl}^{(\tt s)},
  \end{align}
  and for $U_{nl} \in \mathcal{U}_{l}^{\tt m}$,
  \begin{align}\label{mean square gl00 glnl}
  \bbE_\bh\left\{\left|\bg_{l00}^{{(\tt m)}H}\bhg^{(\tt m)}_{lnl}\right|^2\right\}= M_{\tt m}\varphi_{l00}^{(\tt m)}\varphi_{lnl}^{(\tt m)}\eta_{lnl}^{(\tt m)},
  \end{align}
\end{itemize}

Applying all these expectations into \eqref{rate mL jensen} gives
\begin{align}\label{rate mL 1}
&\bbE\left\{\log_2\left(1+{\tt SINR}_{\tt m}^{\tt L}\right)\right\}\notag\\
 &~~~~~\approx \bbE_{\bs\varphi}\left\{\log_2 \left(1+\frac{P_{\tt m} \left(M_{\tt m}+1\right)\varphi_{000}^{{(\tt m)}}\eta_{000}^{(\tt m)}/\left|\mathcal{U}_0^{\tt m}\right|}{P_{\tt m} \sum\limits_{l \in \Phi_{\tt m}} \varphi_{l00}^{(\tt m)} + P_{\tt s} \sum\limits_{j \in \Phi_{\tt s}} \varphi_{j00}^{(\tt s)} - P_{\tt m} \varphi_{000}^{(\tt m)}\eta_{000}^{(\tt m)}/\left|\mathcal{U}_0^{\tt m}\right|+\mu_1 +\sigma^2}\right)\right\},
\end{align}
where
\begin{equation}\label{mu1}
\mu_1 \triangleq \sum\limits_{l \in \Phi_{\tt m} \backslash 0}\left[\frac{\mathds{1}\left(U_{0l} \in \mathcal{U}_l^{\tt m} \right)P_{\tt m} M_{\tt m} \varphi_{l00}^{{(\tt m)}2}}{\left|\mathcal{U}_l^{\tt m}\right|\left(\varphi_{l0l}^{(\tt m)}+\sum\limits_{l' \in \Phi_{\tt m}\backslash l}\varphi_{l0l'}^{(\tt m)}+\frac{\sigma^2}{\tau p_p}\right)} + \frac{\mathds{1}{\left(U_{0l} \in \mathcal{U}_{q_l}^{\tt s} \right)}P_{\tt s} M_{\tt s} \varphi_{q_l00}^{{(\tt s)}2}}{\left|\mathcal{U}_{q_l}^{\tt s}\right|\left(\varphi_{q_l0l}^{(\tt s)}+\sum\limits_{l' \in \Phi_{\tt m}\backslash l}\varphi_{q_l0l'}^{(\tt m)}+\frac{\sigma^2}{\tau p_p}\right)}\right].
\end{equation}

Considering that the path loss from interfering cells are much smaller than that from the associated cell, we approximate the large-scale fading coefficients of interfering cells by their means.
Therefore\footnote{To simplify notation, we use $\bbE$ to denote $\bbE_{\bs \varphi}$ when the expectation is only averaging over the large-scale fading.}, we have
\begin{multline}\label{mu1 approximation}
\mu_1 \approx \tilde \mu_1 = \frac{P_{\tt m} M_{\tt m} \bbE\left\{\sum\limits_{l \in \Phi_{\tt m} \backslash 0}\varphi_{l00}^{{(\tt m)}2}\right\}\frac{\bbE\left\{\mathds{1}{\left(U_{0l} \in \mathcal{U}_l^{\tt m} \right)}\right\}}{\bbE\left\{\left|\mathcal{U}_l^{\tt m}\right|\right\}}}{\bbE\left\{\varphi_{l0l}^{(\tt m)}\right\}+\bbE\left\{\sum\limits_{l' \in \Phi_{\tt m}\backslash l}\varphi_{l0l'}^{(\tt m)}\right\}+\frac{\sigma^2}{\tau p_p}}
 + \frac{P_{\tt s} M_{\tt s}\bbE\left\{ \sum\limits_{l \in \Phi_{\tt m} \backslash 0}\varphi_{q_l00}^{{(\tt s)}2}\right\}\frac{\bbE\left\{\mathds{1}{\left(U_{0l} \in \mathcal{U}_{q_l}^{\tt s} \right)}\right\}}{\bbE\left\{\left|\mathcal{U}_{q_l}^{\tt s}\right|\right\}}}{\bbE\left\{\varphi_{q_l0l}^{(\tt s)}\right\}+\bbE\left\{\sum\limits_{l' \in \Phi_{\tt m}\backslash l}\varphi_{q_l0l'}^{(\tt m)}\right\}+\frac{\sigma^2}{\tau p_p}}.
\end{multline}
\hspace{-4pt}
Next, we try to derive the expectations in \eqref{mu1 approximation}.

For $l \in \Phi_{\tt m} \backslash 0$, when $U_{0l}$ is associated with the MBS $l$ in LoS, we obtain
\begin{equation}\label{mean indicator 1}
\bbE\left\{\mathds{1}{\left(U_{0l} \in \mathcal{U}_l^{\tt m} \right)}\right\}=\bbP\left[E_{\tt m}^{\tt L}\right]=\mathcal{A}_{\tt m}^{\tt L},
\end{equation}
and
\begin{equation}\label{mean v1}
\bbE\left\{\varphi_{l0l}^{(\tt m)}\right\}=\int_0^\infty L^{\tt L} r^{-\alpha^{\tt L}} f_{R_{\tt m}^{\tt L}} (r) dr.
\end{equation}
Similar results can be got for $U_{0l}$ associated with the MBS $l$ in NLoS, associated with the SCB $j$ in LoS and NLoS, respectively.
The average number of association users in \eqref{mu1 approximation} should be calculated conditioned on $U_{0l} \in \mathcal{U}_{l}^{\tt m}$ or $U_{0l} \in \mathcal{U}_j^{\tt m}$. Then, we have
\begin{equation}\label{average number of users}
 \bbE\left\{\left|\mathcal{U}_l^{\tt m}\right|\right\}= \frac{\mathcal{A}_{\tt m}}{\lambda_{\tt m}}\left(\lambda_{\tt m} N -1\right) +1,~~\bbE\left\{\left|\mathcal{U}_j^{\tt s}\right|\right\}= \frac{\mathcal{A}_{\tt s}}{\lambda_{\tt s}}\left(\lambda_{\tt m} N-1\right) +1.
\end{equation}
Considering that the interference from other cells are generally quite small, according to {\it Assumption \ref{assumption 1}}, we can get the following approximation from the Campbell's theorem \cite{baccelli09} as follows:
\begin{equation}\label{rate mean 3}
 \bbE\left\{ \sum\nolimits_{l \in \Phi_{\tt m}\backslash 0 }  \varphi_{l00}^{(\tt m)2} \right\}  \approx 2\pi \lambda_{\tt m} \int_{C_{\tt v}}^\infty r\left[\left(L^{\tt L}\right)^2 r^{-2\alpha^{\tt L}} \bbP_{\tt m}^{\tt L}(r) + \left(L^{\tt NL}\right)^2 r^{-2\alpha^{\tt NL}}\left(1- \bbP_{\tt m}^{\tt L}(r)\right)\right]dr,
\end{equation}
and
\begin{equation}\label{rate mean 4}
 \bbE\left\{\sum\nolimits_{l \in \Phi_{\tt s}\backslash 0}  \varphi_{q_l00}^{(\tt s)2} \right\} \approx 2\pi \lambda_{\tt m} \int_{C_{\tt v}}^\infty r\left[\left(L^{\tt L}\right)^2 r^{-2\alpha^{\tt L}} \bbP_{\tt s}^{\tt L}(r) + \left(L^{\tt NL}\right)^2 r^{-2\alpha^{\tt NL}}\left(1- \bbP_{\tt s}^{\tt L}(r)\right)\right]dr.
\end{equation}
Moreover, according to {\it Assumption \ref{assumption 1}} and {\it \ref{assumption 2}}, for any $l$, the expectation of the path loss from all interfering users are given by
\begin{equation}\label{rate mean 5}
 \bbE\left\{\sum\nolimits_{l' \in \Phi_{\tt m} \backslash l } \varphi_{l 0 l'}\right\} =2 \pi \int_{C_{\tt v}}^\infty u \lambda_{\mathcal{N}'_0}(u) \left[L^{\tt L} u^{-\alpha^{\tt L}} \bbP_{\tt m}^{\tt L} (u) + L^{\tt NL} u^{-\alpha^{\tt NL}}\left(1-\bbP_{\tt m}^{\tt L} (u) \right)\right] du,
\end{equation}
and
\begin{equation}\label{rate mean 6}
 \bbE\left\{\sum\nolimits_{l' \in \Phi_{\tt m} \backslash l } \varphi_{q_l 0 l'}^{(\tt s)} \right\}
\approx 2 \pi \int_{C_{\tt v}}^\infty u \lambda_{\mathcal{N}'_0}(u) \left[L^{\tt L} u^{-\alpha^{\tt L}} \bbP_{\tt s}^{\tt L} (u) + L^{\tt NL} u^{-\alpha^{\tt NL}}\left(1-\bbP_{\tt s}^{\tt L} (u) \right)\right] du.
\end{equation}

The substitution of all these expectations into \eqref{mu1 approximation} gives
\begin{equation}\label{mu1 approximation final}
\tilde \mu_1= \frac{\mathcal{A}_{\tt m}^{\tt L} P_{\tt m} M_{\tt m}\xi_1 }{\mathbb{N}_{\tt m} \left(\nu_1 \nmsp +\nmsp  \chi_1 \nmsp +\nmsp \frac{\sigma^2}{\tau p_p} \right) } +  \frac{\mathcal{A}_{\tt m}^{\tt NL} P_{\tt m} M_{\tt m}\xi_1 }{\mathbb{N}_{\tt m} \left(\nu_2 \nmsp +\nmsp  \chi_1\nmsp  +\nmsp \frac{\sigma^2}{\tau p_p} \right) }
 +  \frac{\mathcal{A}_{\tt s}^{\tt L} P_{\tt s} M_{\tt s}\xi_2  }{\mathbb{N}_{\tt s} \left(\nu_3 \nmsp +\nmsp  \chi_2 \nmsp +\nmsp \frac{\sigma^2}{\tau p_p} \right) }+ \frac{\mathcal{A}_{\tt s}^{\tt NL} P_{\tt s} M_{\tt s}\xi_2  }{\mathbb{N}_{\tt s} \left(\nu_4 \nmsp +\nmsp  \chi_2 \nmsp  +\nmsp \frac{\sigma^2}{\tau p_p} \right) }.
\end{equation}
Therefore, \eqref{rate mL 1} becomes
\begin{align}
\bbE\left\{\log_2\left(1+{\tt SINR}_{\tt m}^{\tt L}\right)\right\}
& \approx \bbE_{\bs\varphi}\nsp \left\{\nsp \log_2 \nsp \left(\nsp 1+\frac{\frac{P_{\tt m}}{\mathbb{N}_{\tt m}} \left(M_{\tt m}+1\right)\varphi_{000}^{{(\tt m)}}\eta_{000}^{(\tt m)}}{P_{\tt m} \sum\limits_{l \in \Phi_{\tt m}} \varphi_{l00}^{(\tt m)} + P_{\tt s} \sum\limits_{j \in \Phi_{\tt s}} \varphi_{j00}^{(\tt s)} - \frac{P_{\tt m}}{\mathbb{N}_{\tt m}} \varphi_{000}^{(\tt m)}\eta_{000}^{(\tt m)}+\rho_1}\right)\nsp \right\}\notag\\
& \approx \bbE_{\bs\varphi}\nsp \left\{\nsp \log_2 \nsp \left(\nsp 1+\frac{\frac{P_{\tt m} \left(M_{\tt m}+1\right)\varphi_{000}^{{(\tt m)}2}}{\mathbb{N}_{\tt m}\left(\varphi_{000}^{(\tt m)}+\chi_1+\sigma^2/\tau p_p\right)}}{P_{\tt m} \nsp \sum\limits_{l \in \Phi_{\tt m}}\nsp \varphi_{l00}^{(\tt m)}\nsp  + \nsp P_{\tt s} \nsp \sum\limits_{j \in \Phi_{\tt s}}\nsp \varphi_{j00}^{(\tt s)} - \frac{P_{\tt m} \varphi_{000}^{(\tt m)2}}{\mathbb{N}_{\tt m}\left(\varphi_{000}^{(\tt m)}+\chi_1+\sigma^2/\tau p_p\right)}\nsp +\nsp \rho_1}\right)\nsp \right\}\label{rate mL 2 temp1}\\
&\mathop = \limits^{(a)} \int_0^\infty \int_0^\infty \frac{e^{-z}}{z\ln 2} \left(\bbE\left\{e^{-z\Delta_1}\right\}-\bbE\left\{e^{-z\Delta_2}\right\}\right)f_{R_{\tt m}^{\tt L}}(r) dz dr,\label{rate mL 2}
\end{align}
where $(a)$ is from the continuous mapping theorem \cite{hamdi10}, and
\begin{equation}\label{Delta 1}
\Delta_1 \triangleq \frac{P_{\tt m}}{\rho_1} \sum\limits_{l \in \Phi_{\tt m} \backslash 0} \varphi_{l00}^{(\tt m)}+ \frac{P_{\tt s}}{\rho_1} \sum\limits_{j \in \Phi_{\tt s}} \varphi_{j00}^{(\tt s)} + \frac{P_{\tt m}}{\rho_1} \left[L^{\tt L} r^{-\alpha^{\tt L}} - \frac{\left(L^{\tt L}\right)^2 r^{-2\alpha^{\tt L}}}{\mathbb{N}_{\tt m}\left(L^{\tt L} r^{-\alpha^{\tt L}}+\chi_1+\sigma^2/\tau p_p\right)} \right],
\end{equation}
\begin{equation}\label{Delta 2}
\Delta_2 \triangleq \frac{P_{\tt m}}{\rho_1} \sum\limits_{l \in \Phi_{\tt m} \backslash 0} \varphi_{l00}^{(\tt m)}+ \frac{P_{\tt s}}{\rho_1} \sum\limits_{j \in \Phi_{\tt s}} \varphi_{j00}^{(\tt s)} + \frac{P_{\tt m}}{\rho_1} \left[L^{\tt L} r^{-\alpha^{\tt L}}+ \frac{M_{\tt m}\left(L^{\tt L}\right)^2 r^{-2\alpha^{\tt L}}}{\mathbb{N}_{\tt m}\left(L^{\tt L} r^{-\alpha^{\tt L}}+\chi_1+\sigma^2/\tau p_p\right)} \right],
\end{equation}
while $\rho_1 \triangleq \tilde \mu_1+ \sigma^2$.
With the probability generating functional of the PPP \cite{stoyan96}, we have
\begin{align}\label{laplace 1}
\hspace{-2ex}&\bbE\left\{\exp\left(-z\frac{P_{\tt m}}{\rho_1} \sum\nolimits_{l \in \Phi_{\tt m}\backslash 0} \varphi_{l00}^{(\tt m)}\right)\right\} \notag\\
 = &\bbE\left\{\exp\left(-z\frac{P_{\tt m}}{\rho_1} \sum\nolimits_{l \in \Phi_{\tt m}\backslash 0} \varphi_{l00}^{(\tt m),\tt L}\right)\right\}\bbE\left\{\exp\left(-z\frac{P_{\tt m}}{\rho_1} \sum\nolimits_{l \in \Phi_{\tt m}\backslash 0} \varphi_{l00}^{(\tt m),\tt NL}\right)\right\}\notag\\
=&\exp\left(\nsp -2 \pi \lambda_{\tt m}\left[ \int_{r}^\infty \nsp  \left(\nsp 1 - e^{-\frac{zP_{\tt m}L^{\tt L}}{\rho_1 u^{\alpha^{\tt L}}} }\right)\bbP_{\tt m}^{\tt L}(u) u du +\nsp \int_{k_1 r^\frac{\alpha^{\tt L}}{\alpha^{\tt NL}}}^\infty\nsp \left(\nsp 1- e^{-\frac{zP_{\tt m}L^{\tt NL}}{\rho_1u^{\alpha^{\tt NL}}} }\right)\nsp  \left(1 \nmsp - \nmsp \bbP_{\tt m}^{\tt L}(u)\right) u du \right]\right),
\end{align}
and
\begin{align}\label{laplace 2}
&\bbE\left\{\exp\left(-z\frac{P_{\tt s}}{\rho_1} \sum\nolimits_{j \in \Phi_{\tt s}} \varphi_{j00}^{(\tt s)}\right)\right\} \notag\\
=& \bbE\left\{\exp\left(-z\frac{P_{\tt s}}{\rho_1} \sum\nolimits_{j \in \Phi_{\tt s}} \varphi_{j00}^{(\tt s),\tt L}\right)\right\}\bbE\left\{\exp\left(-z\frac{P_{\tt s}}{\rho_1} \sum\nolimits_{j \in \Phi_{\tt s}} \varphi_{j00}^{(\tt s),\tt NL}\right)\right\}\notag\\
=&\exp\nsp \left(\nsp -2 \pi \lambda_{\tt s} \nsp \left[\int_{k_2 r}^\infty \nsp \left(\nsp 1-e^{-\frac{zP_{\tt s}L^{\tt L}}{\rho_1u^{\alpha^{\tt L}}} }\right)\bbP_{\tt s}^{\tt L}(u) u du +\nsp \int_{k_1 k_3 r^\frac{\alpha^{\tt L}}{\alpha^{\tt NL}}}^\infty\nsp \left(\nsp 1-e^{-\frac{zP_{\tt s}L^{\tt NL}}{\rho_1u^{\alpha^{\tt NL}}} }\right)\nsp \left( 1 \nmsp - \nmsp \bbP_{\tt s}^{\tt L}(u)\right) u du \right]\right).
\end{align}

The rate $\bbE\left\{\log_2\left(1+{\tt SINR}_{\tt m}^{\tt L}\right)\right\}$ can be obtained by substituting \eqref{Delta 1}--\eqref{laplace 2} into \eqref{rate mL 2}.
Following the same procedure, we can also derive the rate $\bbE\left\{\log_2\left(1+{\tt SINR}_{\tt m}^{\tt NL}\right)\right\}$, $\bbE\left\{\log_2\left(1+{\tt SINR}_{\tt s}^{\tt L}\right)\right\}$, and $\bbE\left\{\log_2\left(1+{\tt SINR}_{\tt s}^{\tt NL}\right)\right\}$, respectively. Then, the desired result can be obtained from \eqref{rate comp}.



\end{document}